\documentclass[twocolumn,
reprint,
superscriptaddress,
amsmath,amssymb,
prb,
floatfix,
]{revtex4-2}

\usepackage{graphicx}
\usepackage{dcolumn}
\usepackage{bm}
\usepackage{xcolor}
\usepackage[english]{babel}
\usepackage{ulem,appendix}

\begin{document}

\title{Magneto-Raman effect in van der Waals layered two-dimensional CrSBr antiferromagnet}

\author{Urszula D. Wdowik}
\affiliation{IT4Innovations, V\v{S}B - Technical University of Ostrava, 17. listopadu 2172/15, 70800 Ostrava, Czech Republic} 

\author{Vaibhav Varade}
\affiliation{Department of Condensed Matter Physics, Faculty of Mathematics and Physics, Charles University, Ke Karlovu 3, 121 16 Prague 2, Czech Republic}

\author{Jana Vejpravová}
\affiliation{Department of Condensed Matter Physics, Faculty of Mathematics and Physics, Charles University, Ke Karlovu 3, 121 16 Prague 2, Czech Republic}

\author{Dominik Legut}
\affiliation{IT4Innovations, V\v{S}B - Technical University of Ostrava, 17. listopadu 2172/15, 70800 Ostrava, Czech Republic}
\affiliation{Department of Condensed Matter Physics, Faculty of Mathematics and Physics, Charles University, Ke Karlovu 3, 121 16 Prague 2, Czech Republic}


\begin{abstract}
Magneto-Raman spectroscopy is applied to study spin-phonon coupling in the layered two-dimensional (2D) van der Waals antiferromagnet CrSBr. We report on the effects of temperature and external magnetic field on Raman-active phonons of $A_g$ symmetry in bulk and one-to-six-layer forms of CrSBr that are reflected by the Raman spectral patterns measured at different configurations of circularly polarized laser beam.  The results of experimental investigations reveal that spin-spin and spin-phonon interactions play a significant role below the  N\'{e}el temperature and are notably stronger in CrSBr monolayer than in bulk material. Spin-phonon coupling leads to the emergence of \textit{ new} phonon peaks at very low temperatures. Such effect is solely observed in one-to-six-layer CrSBr and is absent in its bulk form.  Our experimental research is accompanied by \textit{ab initio} simulations of the Raman spectra of bulk and monolayer CrSBr that uncover correlations between intensities of Raman-active phonons and the arrangement of spin magnetic moments on Cr atoms. Comparative analysis of simulated and experimental Raman spectra suggests the most favored spin orientation in the CrSBr antiferromagnetic system.  
\end{abstract}

\maketitle

\section{Introduction}
Van der Waals (vdW) layered two-dimensional (2D) magnetic materials have gained considerable experimental and theoretical interest \cite{Burch2018,Huang2017,Gong2017,Tu2022,Sivadas2018,Huang2018}, not only because of the fascinating physics governing their diverse physical properties but also due to potential application of these materials in the next generation of electronic, optoelectronic and spintronic devices \cite{Wang2012,Lin2018,Shim2017,Liu2017,Tan2020}. Among the 2D magnetic materials with unique megnetoelectronic coupling effects and the possibility to enlarge functionalities of 2D vdW materials, which have been intensively studied during the last few years, is the semiconducting antiferromagnet CrSBr \cite{guo18,Telford2020,Lee2021,yang21,Wilson21,Xu2022,Klein2022,lopez22,Klein23a,Klein23b,pawbake23,Linhart2023,Torres2023,Tschudin2024}. The CrSBr compound belongs to ternary chromium chalcogenide halide compounds (CrXH, where X = S, Se, and H = Cl, Br, and I). Its lattice consists of vdW layers made of two buckled CrS planes terminated by Br atoms (Fig. \ref{fig:structure}). The layers stacking along the crystallographic \textit{c} axis produce an orthorhombic structure with the $Pmmn$ symmetry (space group No. 59, point group $D_{2h}$). Each CrSBr layer orders ferromagnetically (FM) and couples antiferromagnetically (AF) to adjacent layers, maintaining the A-type AF magnetic order below the  N\'{e}el temperature $T_N \sim 132$ K \cite{Telford2020}. The spin magnetic moment on the Cr atom is quite large and amounts to $\sim 3 \mu_{\mathrm{B}}$ \cite{Telford2020,Wilson21,yang21}. CrSBr exhibits triaxial magnetic anisotropy with easy, intermediate and hard magnetization axes along crystallographic \textit{b}, \textit{a}, and \textit{c}  axes, respectively \cite{Goser1990}. It is a semiconducting material with a direct energy gap ranging between 1.25 and 1.5 eV \cite{Telford2020,Wilson21,yang21}. The stability of CrSBr in air and its electronic, optical, and magnetic properties make this compound a good candidate for practical engineering of nano-spintronic, nano-spinphotonic, and memory devices \cite{Klein2022,Klein23a,Klein23b}. 

Along with the upsurge in research in the field of 2D magnetic materials, magneto-Raman spectroscopy, the technique that combines typical Raman spectroscopy with the application of an external magnetic field, has been adopted and widely applied to characterize the magnetic properties of 2D magnets and spin-phonon interaction \cite{Lee2016,Kim2019a,Kim2019b,Sun2021,Zhang2020,McCreary2020a,McCreary2020b,Kong2024,Staros2022}. 

This paper addresses a magneto-Raman examination of the impact of external magnetic field and temperature on the coupling between spin magnetic moments on the chromium sublattice and atomic vibrations in this 2D semiconducting vdW-layerd CrSBr antiferromagnet and explores this effect as a function of number of material layers (from monolayer to bulk). The present experiments are supported by first-principles calculations of the phonon dynamics and Raman spectra in bulk and monolayer CrSBr with various alignment of spin magnetic moments within the A-type AF order.          

\section{Methodology\label{methods}}
\subsection{Experimental}

High-quality CrSBr crystals were synthesized by chemical vapor transport (CVT) by treating a stoichiometric ratio of three corresponding elements, namely chromium, sulfur, and bromine, in a temperature gradient in a sealed quartz tube under high vacuum. Initially, the temperature at both ends was maintained between 300\textsuperscript{o}C and 700\textsuperscript{o}C, which was subsequently increased to 800\textsuperscript{o}C and 900\textsuperscript{o}C while altering the ends. The reaction was continued over 10 days and high purity CrSBr crystals were obtained. Different structures of atomically thin monolayers and other layers were exfoliated manually using scotch tape and transferred onto a marked substrate fabricated using photolithography. Using Bruker's AFM Dimension ICON system in the quantitative non-mechanical mode, we captured AFM images and thickness profiles of the structures. The AFM data were acquired with a Bruker silicon tip, and the subsequent processing and analysis of the AFM images were carried out using Gwyddion software. Layer-dependent room-temperature Raman spectra of CrSBr and CrSBr were collected in the backscattering geometry with an excitation wavelength of 532 nm (2.33 eV) at 200 $\mu$ W power using a WITec spectrometer in the ambient environment using a grating of 1800 lines/mm, respectively. 

The circularly polarized cryomagnetic Raman spectra were measured in the back-scattering geometry using a low-temperature confocal Raman microscope insert (attoRAMAN, attocube) placed in a Physical Property Measurement System (PPMS, Quantum Design). We utilized a 100 $\times$ objective lens compatible with low temperatures and magnetic fields, featuring a numerical aperture of 0.82 and a lateral resolution of 500 nm. This lens was used to focus the 532 nm laser beam with intensity 500 $\mu$. The acquisition of low-frequency Raman spectra was extended down to 10 cm$^{-1}$ by incorporating two volume Bragg grating notch filters into the Raman system, effectively mitigating the Rayleigh signal. The circular polarization of the incident laser beam by employing a conventional set of half- and quarter-wave plates designed for a 532 nm laser. Similarly, an array of broadband quarter and half-wave plates was used to obtain polarization-resolved emitted signals. The degree of polarization was precalibrated for each incident and scattered beam, with an ellipticity exceeding 44.5$^{o}$, using ThorLabs' TXP polarimeter. The spectrometer, equipped with an 1800 lines/cm grating, achieved a spectral resolution of 0.6 cm$^{-1}$ under the measurement conditions. The CCD detector's intensity response was precalibrated using a tungsten halogen light source (HL-2000-CAL, Ocean Optics). In addition to the PPMS temperature readout, the actual sample temperature was monitored with a thermometer located on the optical insert at the sample position. Spectra were recorded during the cooling cycle from 300 to 10 K, with data acquisition occurring after temperature stabilization at the measured point. The magnetic field evolution of Raman spectra was obtained using time series started manually with the magnetic field sequence in PPMS.

\subsection{Theoretical}
Our theoretical studies were carried out within the framework of spin-polarized density functional theory (DFT) implemented in the plane-wave basis VASP code \cite{vasp1,vasp2}. The Perdew-Burke-Ernzerhof (PBE) generalized gradient approximation (GGA) exchange-correlation functional \cite{pbe1} and the projector-augmented wave (PAW) method \cite{paw1,paw2} were employed for the description of the electron-ion interactions. Calculations were performed with an energy cut-off for the plane wave expansion of 340~eV. The configurations Cr($3d^5$$4s^1$), S($3s^2$$3p^4$) and Br($4s^2$$4p^5$) were treated as valence electrons. Strong local interactions on site in the $3d$ shell of Cr were taken into account via the simplified rotationally invariant approach \cite{dudarev98} with an effective Hubbard potential of 3~eV \cite{guo18,yang21,pawbake23}. The semi-empirical corrections of Grimme et al. with Becke-Johnson damping (DFT-D3) \cite{vdw1,vdw2} were incorporated to account for van der Waals interactions. The bulk CrSBr structures of A-type antiferromagnetic (AF) with ferromagnetically (FM) aligned spins along $b$-axis (magnetic easy axis), the $a$ axis (magnetic intermediate axis) and the $c$ axis (magnetic hard axis) were fully optimized using the 8x8x4 Monkhorst-Pack grid of $k$ points. The resulting theoretical structures are consistent with those reported in recent experimental studies \cite{lopez22,Lee2021} (see Table \ref{tab:structure} in Appendix \ref{appendix}). In addition, we took into account the CrSBr monolayer (ML), which was created from the optimized bulk material by adding a vacuum space of approximately 15~\AA~ along the direction perpendicular to a layer to avoid interactions between the CrSBr sheet and its periodic images.  ML structures with spins along the $a$ -, $b$ - and $c$ axes were optimized with 8x8x1 $k$ point meshes and convergence criteria for the total energy and internal forces of $10^{-8}$~eV and $10^{-6}$~eV~\AA$^{-1}$, respectively.

The phonons of the modeled structures were determined within the harmonic approximation and using a direct method \cite{parlinski97,togo15}, which utilizes the Hellmann-Feynman forces calculated by DFT acting on all atoms in a given supercell. The 4x4x2 and 4x4x1 supercells for bulk and ML were applied. Symmetry non-equivalent atoms were displaced from their equilibrium positions with the amplitude of 0.01~\AA. The frequencies and polarization vectors of the phonons were obtained by solving an eigenvalue problem for the respective dynamical matrices. The peak intensities of the nonresonant Raman spectra (in the Stokes process) were calculated according to the expression \cite{cardona83}: $I \propto |\mathbf{e}_\mathrm{i} \mathbf{R} \mathbf{e}_\mathrm{s}|^2 \omega^{-1} (n+1)$, where $(n+1)$ is the population factor for Stokes scattering with $n=[ \exp(\hbar/k_BT)^{-1}]^{-1}$ denoting the Bose-Einstein thermal factor, $\mathbf{e}_\mathrm{i}$ ($\mathbf{e}_\mathrm{s}$) is the polarization of incident (scattered) radiation, $\mathbf{R}$ is the Raman susceptibility tensor. The components of the $\mathbf{R}$ tensor were determined from derivatives of the electric polarizability tensor over the atomic displacements \cite{umari01,umari03}.


\section{Results and discussion\label{results}}
\begin{figure}
    \centering
    \includegraphics[width=0.95\linewidth]{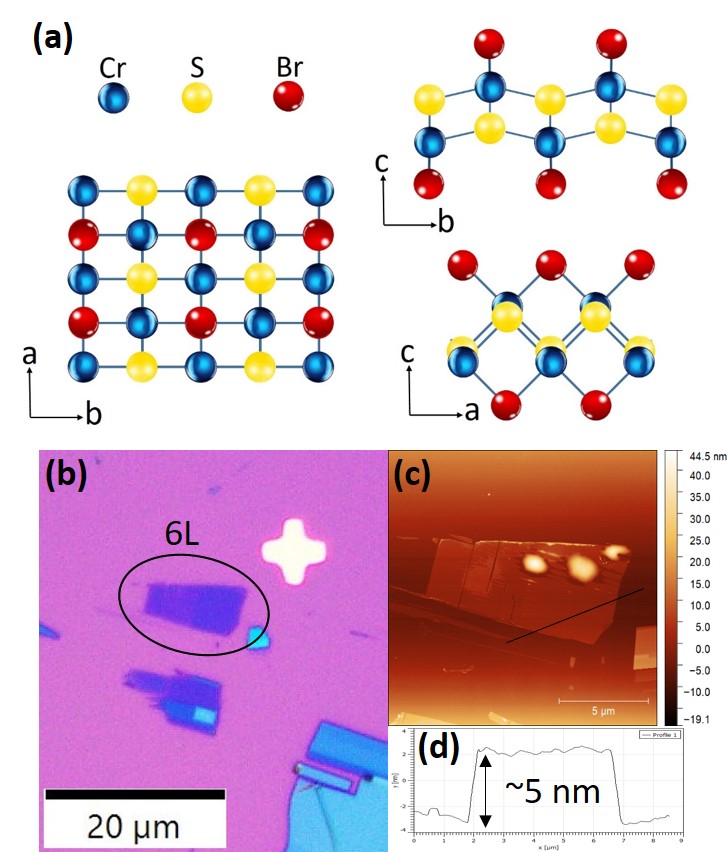}
    \caption{(a) Schematic lattice structure of CrSBr along a, b, and c axes. (b) Optical microscope image of 6L (marked by circle) and bulk CrSBr, (c) Atomic force microscopy (AFM) image, and (d) thickness profile measured along the path indicated by black line.}
    \label{fig:structure}
\end{figure}

The crystal structure of the CrSBr vdW magnet along three crystallographic axes is schematically shown in Fig.~\ref{fig:structure}(a). In the \textit{cb} plane, {\it i.e.}, along the \textit{a}-axis, each single layer of the CrSBr crystal is composed of two knotted planes constituted by Cr and S atoms that are sandwiched between Br atomic sheets. The layers are stacked in the \textit{ab} plane along the \textit{c} axis with different interplanar spacing along the \textit{a} and \textit{b} axes \cite{Telford2020}. Figures~\ref{fig:structure}(b)-(d) show a microscope image of 6 layers (6L) of CrSBr on which most of our measurements were performed, its AFM image, and its thickness profile, respectively, while the microscope and AFM images of 1L, 2L, 3L, 5L, and bulk CrSBr with a number of layers greater than 70 are displayed in Fig.~\ref{fig:AFM&profiles} (Appendix \ref{appendix}). Each layer exhibits in-plane FM order. The FM layers are coupled antiferromagnetically along the stacking direction. The crystal and magnetic structure features of CrSBr suggest that the effect of spin-phonon coupling in this layered system may be notably significant \cite{pawbake23}.   
Therefore, to investigate the impact of spin polarization on phonon dynamics in bulk and monolayer CrSBr, especially on their Raman spectra, \textit{ab inito} simulations were undertaken in the first stage. The resulting phonon spectra for the spin arrangements AF-a, AF-b, and AF-c, which are shown in Fig. ~\ref{fig:phDOS} of Appednix \ref{appendix} clearly indicate a negligible effect of the spin polarization on the phonon spectrum of both bulk and ML CrSBr.  This is mainly due to the insignificant differences in interatomic distances between the AF-a, AF-b, and AF-c configurations considered, as evidenced by the lack of visible changes between their structural parameters (Table \ref{tab:structure}). Very weak spin-lattice coupling leads to almost identical force constants which govern the phonon dynamics in these configurations. However, remarkable changes are observed between the phonon spectra of the bulk and ML forms. The most significant differences between the bulk and ML phonon spectra are found for their lowest ($\omega<80$ cm$^{-1}$) and highest ($\omega > 280$ cm$^{1-}$) frequency ranges, for which softening and hardening of modes are observed, respectively, while going from bulk to ML structure. Thus, the Brillouin zone center phonon modes, i.e., the Raman- and infrared-active ones, are expected to follow a similar trend. 

The analysis of $\Gamma$-point phonons, Raman tensors, phonon activities, and the selection rules for CrSBr in bulk and ML forms is based on irreducible representations of the point group $D_{2h}$, and the most relevant information is contained in Table \ref{tab:irreps}, Appendix \ref{appendix}. In general, there are nine Raman-active and nine IR-active phonon modes (including three lattice translational modes) in this system. The three sublattices, i.e., Cr, S, and Br contribute to the Raman-active $A_g$, $B_{2g}$, and $B_{3g}$ phonon modes. The atomic displacements associated with particular Raman modes are schematically depicted in Fig.~\ref{fig:atomic_displacements}, Appendix \ref{appendix}.  The IR-active phonons with $B_{1u}$, $B_{2u}$, and $B_{3u}$ symmetries correspond to the oscillations of the dipole moment parallel to the crystal \textit{c}-, \textit{b}-, and \textit{a}-axis, respectively. These considerations apply to bulk as well as monolayer CrSBr. 

\begin{figure}
    \centering
    \includegraphics[width=0.95\linewidth]{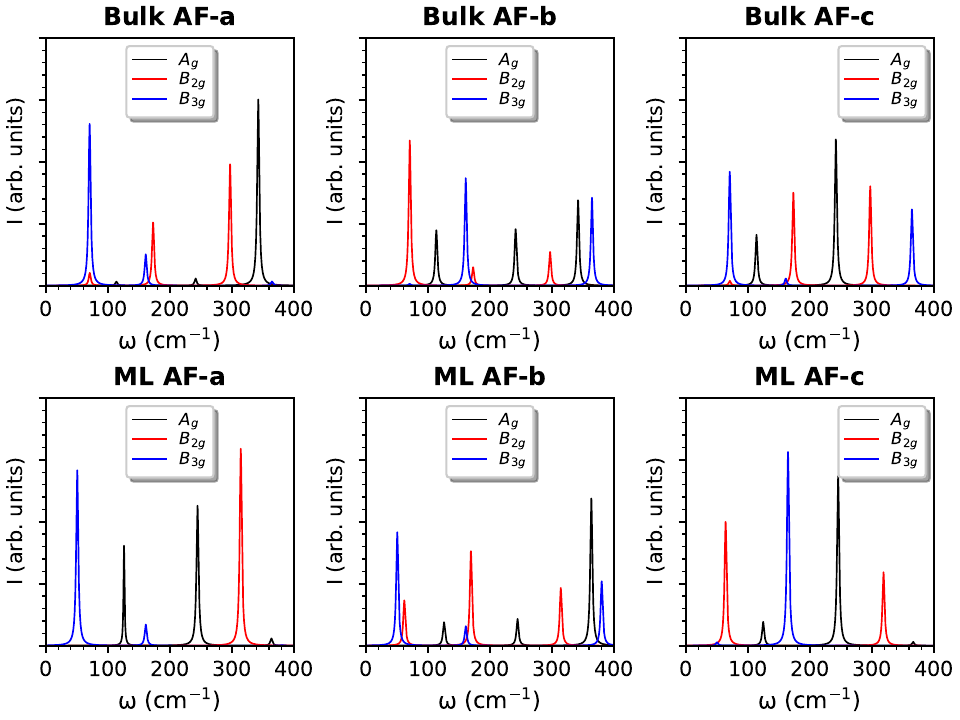}
    \caption{Linearly polarized backscattering Raman spectra for bulk and ML CrSBr with AF-a, AF-b and AF-c spin arrangements simulated at 4~K, with the excitation wavelength of 532 nm, and the Lorentzians of artificial HWHMs of 2 cm$^{-1}$. The scattering geometry of $\bar{x}(yy)x$ for $A_g$ modes, $\bar{y}(xz)y$ for $B_{2g}$ modes, and $\bar{x}(yz)x$ for $B_{3g}$ modes is applied.}   
    \label{fig:Ramans_DFT}
\end{figure}

\begin{figure}
    \centering
    \includegraphics[width=0.95\linewidth]{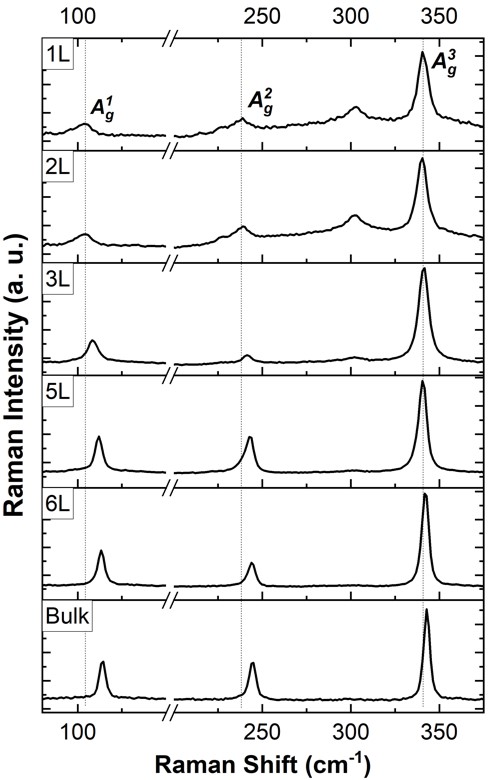}
    \caption{Evolution of the Raman spectra as a function of number of the CrSBr layers.}
    \label{fig:Raman_vs_thickness}
\end{figure}

The resulting frequencies of the phonon modes in the center of the Brillouin zone for the bulk and ML of CrSBr calculated with different spatial spin arrangements are gathered and compared to those determined in the present experiments and other experimental studies \cite{pawbake23,Klein23b} in Table \ref{tab:Gamma_phonons}, \ref{appendix}. Our experimental Raman spectra from both the bulk and 1L systems show three strong $A_g$ modes that are denoted according to their increasing frequency as $A_g^1$, $A_g^2$, and $A_g^3$. In bulk CrSBr, the theoretically predicted frequencies of the $A_g$ modes remain in close correspondence with those obtained in experiments, contrary to its ML form, where they remain shifted to higher frequencies in comparison with those resulting from the present Raman experiments. The lack of a clear dependence of the phonon frequency on the spin orientation makes unambiguous identification of the preferred spin arrangement in CrSBr impossible, when a comparison of the calculated phonon frequencies and the positions of the measured Raman peaks is a sole criterion. However, magnetic interactions can substantially affect not only the position but also the intensities and widths of the Raman active phonons \cite{sternik18}. Here we limit our theoretical considerations to harmonic phonons, and hence we explore only the changes in the intensities of the Raman-active modes owing to particular spin arrangements. Indeed, our theoretical studies indicate significant differences in the intensity of Raman active phonons, not only between ML and bulk CrSBr but also between structures with different spin orientations in the bulk or monolayer material itself. This strong correlation between the intensities of Raman-active phonons and spin arrangement is clearly visible in the simulated linearly polarized Raman spectra depicted in Fig.~\ref{fig:Ramans_DFT}. Although the experimental intensities of the $A
g$ modes are not exactly reproduced by the present computations, a general trend is maintained for both bulk and ML AF-b CrSBr, namely, the $A_g^1$ and $A_g^2$ phonons have much lower intensities than the mode of $A_g^3$ symmetry (cf.   Fig.~\ref{fig:Raman_vs_thickness}). Based on the comparison between the simulated and measured Raman spectra, we may conclude that the spin configuration preferred in CrSBr corresponds to the AF-b, i.e., along the easy magnetization axis.


In addition, the Raman spectra measured as a function of the thickness of the CrSBr layer, Fig.~\ref{fig:Raman_vs_thickness}, clearly show a blueshift of the peaks arising from the $A_g^1$ and $A_g^2$ phonons with an increasing number of layers. However, such a shift is hardly observed for the peak related to the ${A}^{3}_{g}$ mode. Similar behavior has been encountered before \cite{Telford2020}, and is now confirmed by our DFT calculations.

\begin{figure}
    \centering
    \includegraphics[width=0.95\linewidth]{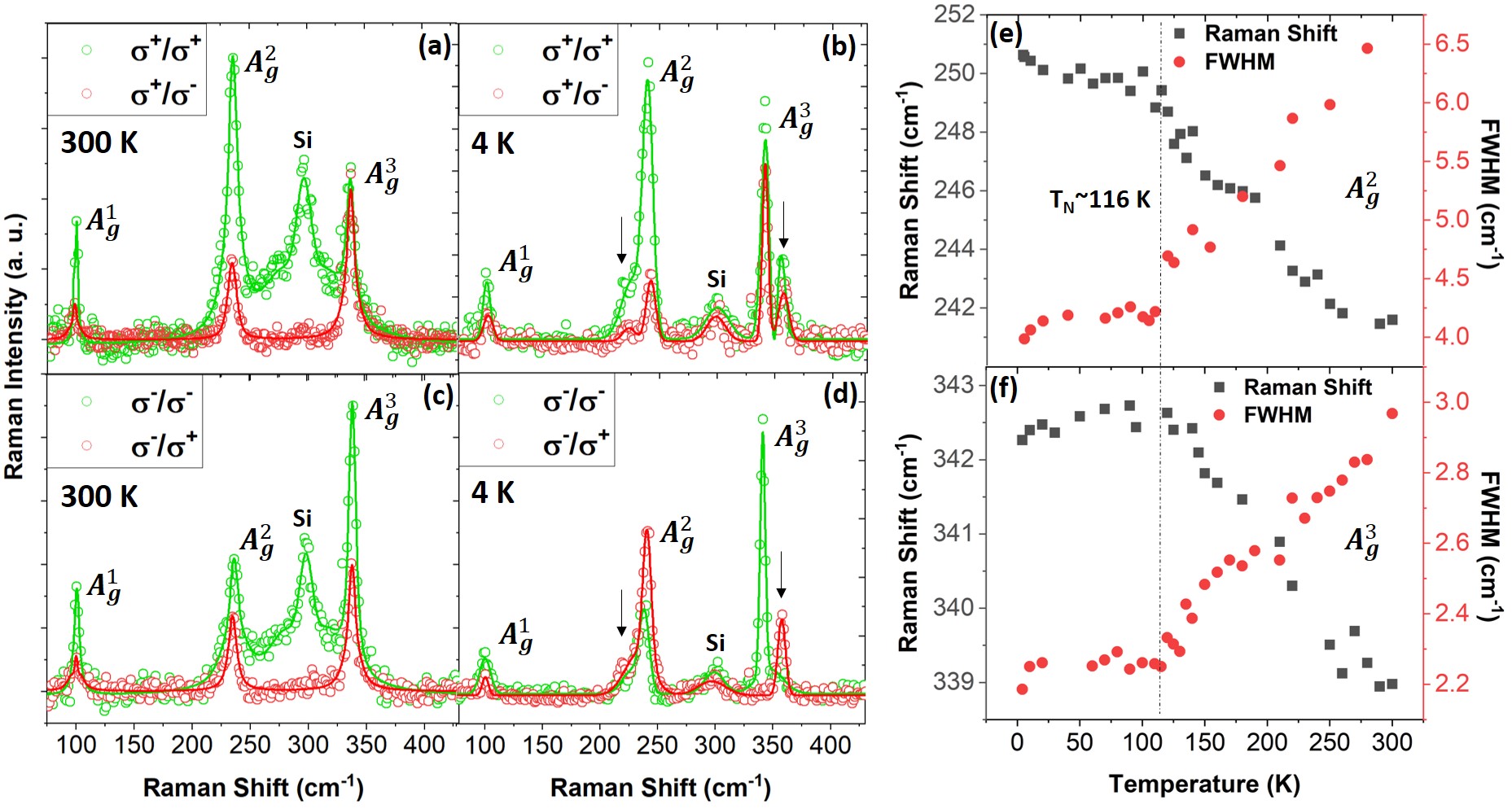}
    \caption{Experimental Raman spectra of 1L CrSBr at 300~K (a, c) and 4~K (b, d) for parallel ($\sigma^{+}/\sigma^{+}$, $\sigma^{-}/\sigma^{-}$) and crossed ($\sigma^{+}/\sigma^{-}$, $\sigma^{-}/\sigma^{+}$) configurations of excitation/detection direction of circularly polarized laser beam of 532~nm wavelength. The symbols $\sigma^{+}$ and and $\sigma^{-}$ state for right and left circular polarizations, respectively. (e, f) Temperature evolution of Raman shifts and widths (FWHM) for $A_g^2$ and $A_g^3$ modes at crossed $\sigma^{+}$/$\sigma^{-}$ configuration.}
    \label{fig:Ram_exp_field_0}
\end{figure}

\begin{figure}
    \centering
    \includegraphics[width=0.9\linewidth]{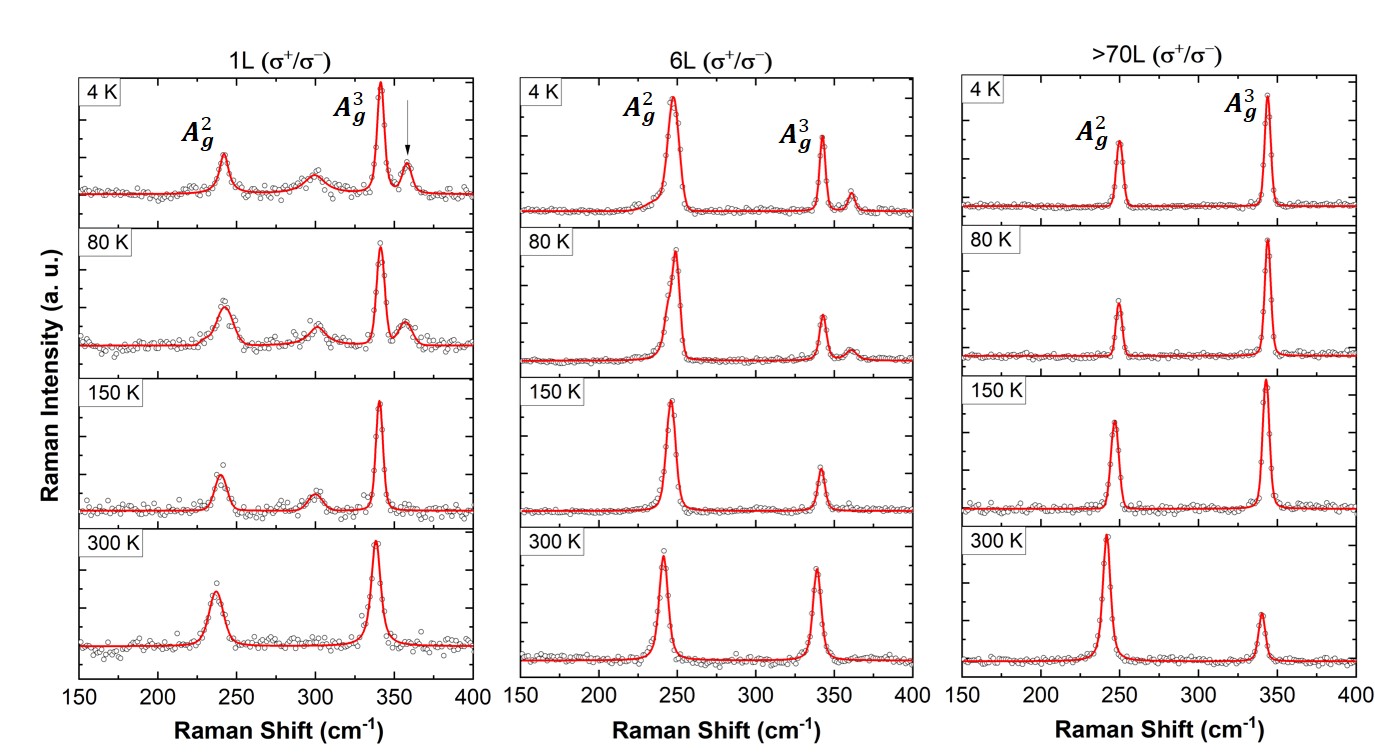}
    \caption{Temperature evolution of the circularly-polarized Raman spectra in 1L, 6L, and bulk ($>70$L) CrSBr at crossed $\sigma^{+}/\sigma^{-}$ polarization configuration.}
    \label{fig:Raman_new_modes}
\end{figure}

A strong axial dependence of the intensities of Raman peaks measured in the linear polarization mode \cite{Telford2020} does not take place for circular polarization, and hence the position of the sample against the direction of laser excitation is irrelevant. However, the intensities of Raman peaks remain sensitive to the right ($\sigma^{+}$) and/or left ($\sigma^{-}$) circular polarizations, as indicated by the circularly-resolved Raman spectra of 1L CrSBr determined at 4 and 300~K for various combinations of excitation/detection combinations that are shown in Fig.~\ref{fig:Ram_exp_field_0}. The influence of circular beam polarization on the modes' intensities is especially noticeable for the $A_g^2$ and $A_g^3$ phonons observed in the parallel configurations ($\sigma^{+}/\sigma^{+}$ and $\sigma^{-}/\sigma^{-}$) at 300 K, which change (revert) their intensities while changing from right ($\sigma^{+}/\sigma^{+}$) to the left ($\sigma^{-}/\sigma^{-}$) polarization and vice versa. This effect practically does not occur at the crossed $\sigma^{+}/\sigma^{-}$ and $\sigma^{-}/\sigma^{+}$ circular polarizations.  
A significant difference in the intensities of both the $A_g^2$ and the $A_g^3$ modes can also be observed in 1L CrSBr at 4~K. Apart from changes in the intensities of Raman-active phonons being observed for both parallel and crossed polarizations, the Raman spectra reveal two additional peaks, as can be seen in Fig.~\ref{fig:Raman_new_modes}. 
They appear close to the $A_g^2$ and $A_g^3$ peaks, on their left and right sides, respectively. The shoulder of the $A_g^2$ peak, which makes the main peak asymmetric, is located at $\sim 220$ cm$^{\mathrm{-1}}$. The second additional peak at $\sim 355$ cm$^{\mathrm{-1}}$ which is adjacent to the $A_g^3$ peak can be assigned to the mode of the $B_g^3$ symmetry. Interestingly, these phonon modes emerge in the spectra only below the N\'{e}el temperature of CrSBr ($T_{\textit{N}} \sim 116$~K), i.e., with the onset of magnetic ordering and/or possible formation of magnetic domains with specific orientation of spins. Moreover, they appear only in the 1L to 6L layers of CrSBr, and they are absent in the bulk system. Enhanced activation of these modes below $T_{\textit{N}}$ can be seen in Fig.~\ref{fig:Raman_new_modes}. The ordered magnetic moments may be responsible for anisotropic effects and distortions in the lattice that will be reflected by the modified intensities, widths, and positions of the Raman peaks. Furthermore, the phonon peaks may experience shifts along with changes in their widths upon temperature due to other reasons, e.g., anharmonicity \cite{calizo07,srivastava18}. The evolution of the Raman peaks' shifts and widths with temperature can be used to analyze the spin-phonon coupling mechanism and to study structural and magnetic phase transitions in bulk and 2D systems \cite{tian16,ghosh21}. In the magnetically ordered CrSBr below $T_{\textit{N}}$ one usually encounters an interplay between spin-spin interactions and lattice vibrations as the propagating phonons affect exchange coupling between magnetic moments \cite{vaclavkova20}. 

The Raman scattering experiments allowed us to identify temperature regions with distinct behavior of the $A_g^2$ and $A_g^3$ phonon modes in the 1L CrSBr. Above $T_{\textit{N}}$, a decrease in the Raman shifts of both modes is assisted by their significant broadening, with an increase in the FWHM of the $A_g^2$ phonon exceeding that for the $A_g^3$ one, as displayed in Figs.~\ref{fig:Ram_exp_field_0} (e)-(f). Such broadening of phonon peaks is likely to be due to enhancement in multiphonon processes and phonon anharmonicity with increasing temperature, which dominate over the effects resulting from possible spin-fluctuations. In the temperature range extending from about 25~K to $T_{\textit{N}}$ a decrease in frequency is observed for the $A_g^2$ phonon, while the frequency of $A_g^3$ undergoes a small increase. In addition, the widths of both phonons do not exhibit essential changes within this temperature range, i.e., they show only some fluctuations around the respective average values acquired at $T_{\textit{N}}$. In this temperature range, spin fluctuations can play a dominant role \cite{Telford2020,Klein23a}. They can gradually decrease with decreasing temperature and finally below some temperature T$^{*}$ maintains a state with frozen spins and possible magnetic domains. Indeed, based on very recent experimental observations, such a process has been proposed and the formation of a new magnetic phase in bulk CrSBr has been suggested as well \cite{lopez22}. This new low-temperature phase in bulk CrSBr was found to stabilize below T$^{*} \sim 40$ K; however, its magnetic and crystal structures are still being debated. Similar observations following from Raman scattering experiments have also been reported for bulk CrSBr, but this new low-temperature phase was observed below T$^{*} \sim 25$~K \cite{pawbake23}. The results of the present studies reveal that the transition to this new low temperature phase can also take place in the 1L CrSBr. Likewise in previous Raman experiments on the bulk system, the transition in 1L CrSBr is found in the vicinity of 25~K, as confirmed by the decrease (increase) in the Raman shift of the peak $A_g^2$ ($A_g^3$) and the slight increase in widths of both phonon modes within the temperature ranging from 4 to 25~K. 

\begin{figure}
    \centering
    \includegraphics[width=0.9\linewidth]{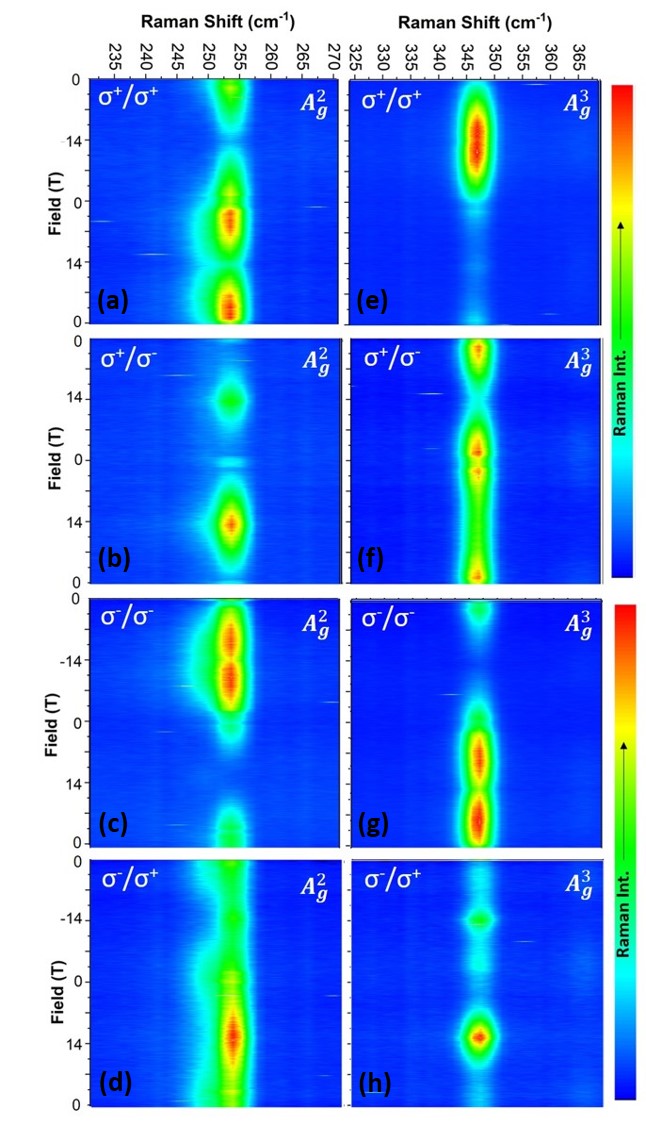}
    \caption{Contour maps of the Raman intensity vs. magnetic field and Raman shift frequency for $A_g^2$ and $A_g^3$ phonon modes of 6L CrSBr in four circular polarization configurations. Mode $A_g^2$: (a) $\sigma^{+}$/$\sigma^{+}$, (b) $\sigma^{+}$/$\sigma^{-}$, (c) $\sigma^{-}$/$\sigma^{-}$, and (d) $\sigma^{-}$/$\sigma^{+}$. Mode $A_g^3$: (e) $\sigma^{+}$/$\sigma^{+}$, (f) $\sigma^{+}$/$\sigma^{-}$, (g) $\sigma^{-}$/$\sigma^{-}$, (h) $\sigma^{-}$/$\sigma^{+}$.}
\label{fig:MagneticFieldRaman}
\end{figure}

\begin{figure}
    \centering
    \includegraphics[width=0.95\linewidth]{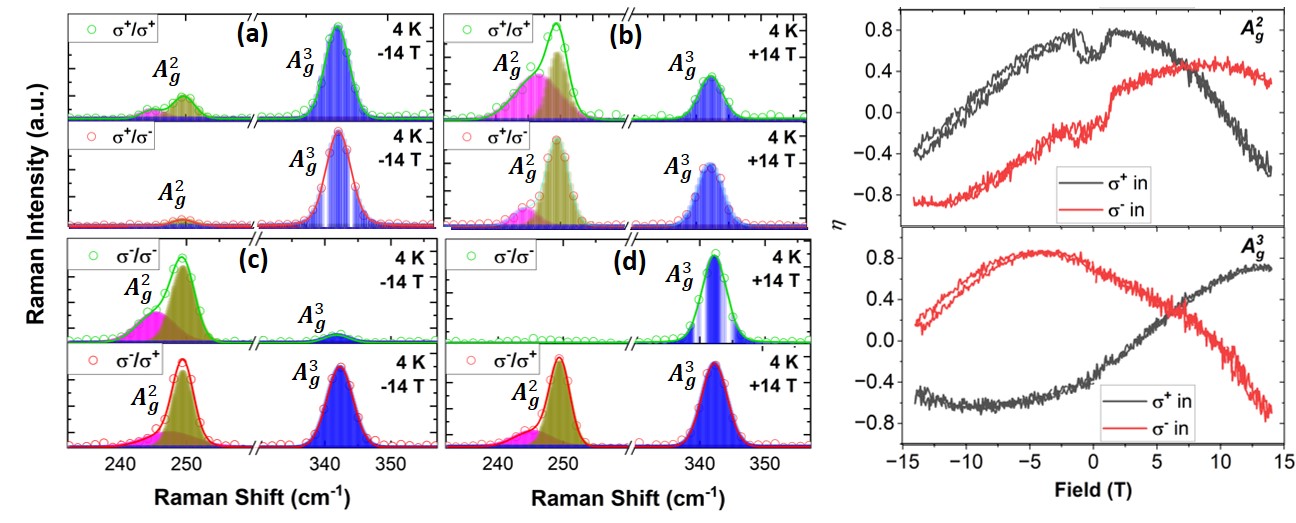}
    \caption{Circular polarization resolved Raman spectra of 6L CrSBr at 4 K in both parallel and cross circular polarization configurations with $\sigma^{+}$ incident at magnetic field (a) -14 T and (b) +14 T, and with $\sigma^{-}$ incident at magnetic field (c) -14 T and (d) +14 T. The (e) and (f) show the degree of polarization ($\eta$) estimated for $A_g^2$ and $A_g^3$ modes for both $\sigma^{+}$ and $\sigma^{-}$ incident. The peaks are fitted with Voigt profile.}
\label{fig:F4}
\end{figure}

Figure \ref{fig:MagneticFieldRaman} illustrates the contour maps of the Raman intensity determined at 4 K for two prominent $A_g^2$ and $A_g^3$ Raman modes in 6L CrSBr as a function of the magnetic field swept between -14 and +14 T. The measurements were carried out in parallel and cross detection modes for the following distinct configurations of circular polarization: $\sigma^{+}$/$\sigma^{+}$, $\sigma^{+}$/$\sigma^{-}$, $\sigma^{-}$/$\sigma^{-}$, and $\sigma^{-}$/$\sigma^{+}$). 
 The variations of the measured Raman modes in parallel and cross-detection are opposite to each other, as expected. The intensity maps reveal the magnetic phase-induced peak \textit{S}, which emerges in close proximity of the mode with $A_g^2$ symmetry. When excited with $\sigma^{+}$, the $A_g^2$ mode gradually diminishes with an increase in the negative magnetic field and almost disappears around -14 T; however, this mode and the \textit{S} peak are found to be most intense around +7 T and 0 T. For cross detection, the mode of $A_g^2$ symmetry decreases significantly around 0 and -7 T while showing a prominent feature around +14 T. Interestingly, the \textit{S} peak remains insignificant in cross-detection and $\sigma^{+}$ excitation. On the other hand, for $\sigma^{-}$ excitation, the $A_g^2$ mode along with the \textit{S} peak are simultaneously prominent in negative magnetic field in parallel detection, while in cross detection, $A_g^2$ is prominent around +14 T and the \textit{S} peak becomes prominent around 0 T. The dependence of the $A_g^2$ phonon and the peak $S$ on the particular configuration of circular polarization in magnetic fields of -14 and +14 at 4~K is illustrated in Fig.~\ref{fig:F4}. It is evident that the peak \textit{S} is not associated with the broadening of the $A_g^2$ phonon mode, but its origin is different and remains related to the magnetic phase of a few CrSBr layers.

According to the evolution of the ${A}^{3}_{g}$ mode as a function of the magnetic field, which is shown in Fig. \ref{fig:MagneticFieldRaman}(e-h), one finds this mode to be prominent in negative magnetic fields, with peaking at -14 T when excited by $\sigma^{+}$, while it is hardly visible in a positive field scan. In cross detection, a gradual decrease is observed near -14 T, whereas it remains rather prominent in other magnetic-field scanning regions. In parallel detection mode, $\sigma^{-}$ excitation, and the magnetic field ranging from 0 to -14 T, the $A_g^3$ mode completely disappears around -14 T, but in the region of positive magnetic fields, this mode becomes quite prominent throughout. It is also intense in close proximity to +14 T in cross-detection mode, but again becomes weak in the remaining magnetic field range. 

\begin{figure}
    \centering
    \includegraphics[width=0.9\linewidth]{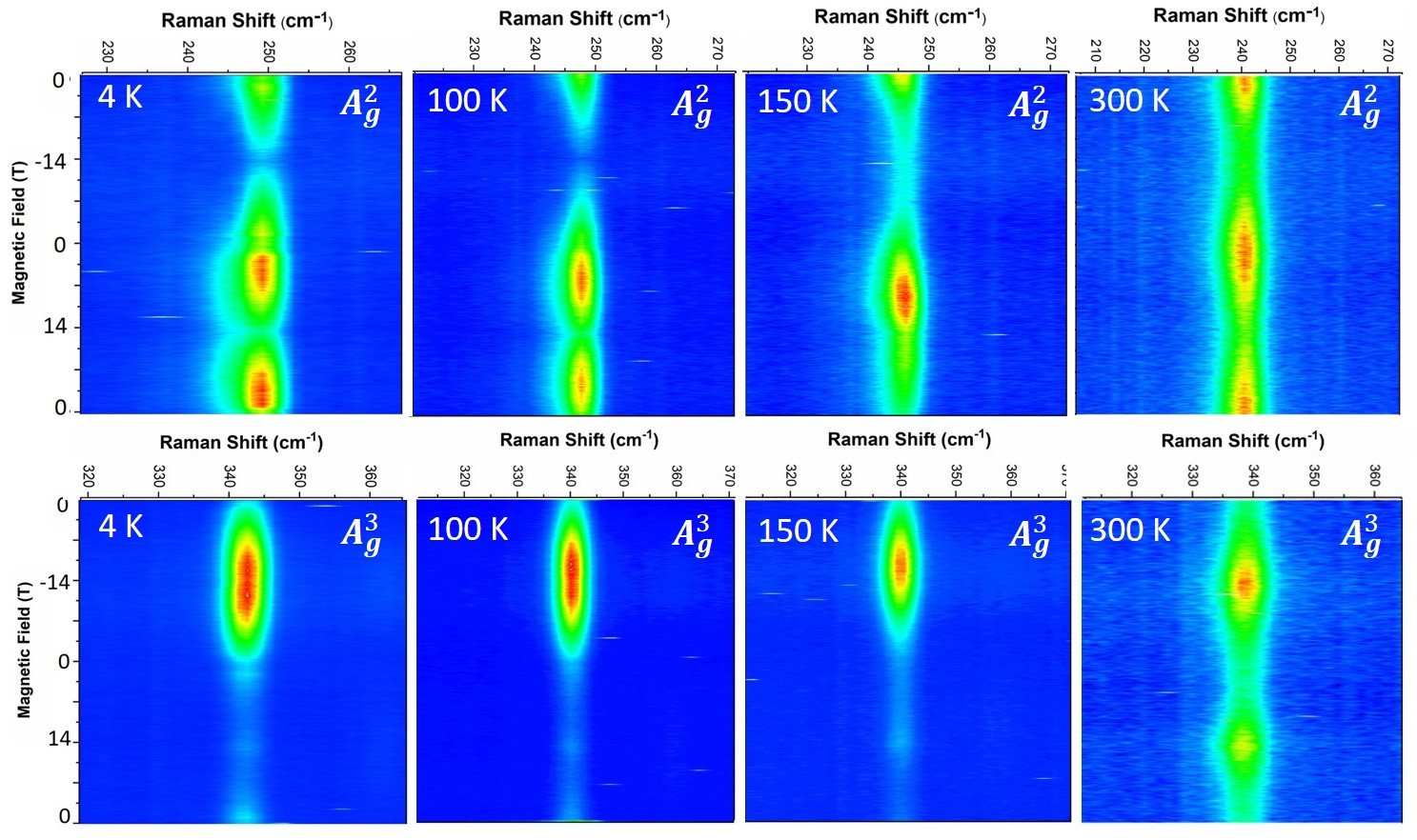}
    \caption{Contour maps of the Raman intensity vs. magnetic field and Raman shift frequency for $A_g^2$ and $A_g^3$ phonon modes of 6L CrSBr at 4, 100, 150, and 300 K in $\sigma^{+}$/$\sigma^{+}$ parallel detection mode.}
\label{fig:FS4}
\end{figure}

To correlate the evolution of Raman modes with the magnetic phase transition, the measurements were also performed as a function of temperature. A comparative evolution of the $A_g^2$ and $A_g^3$ phonon modes for 6L CrSBr in the $\sigma^{+}$/$\sigma^{+}$ parallel detection mode is illustrated in Fig.~\ref{fig:FS4}. The variation in both $A_g^2$ and $A_g^3$ Raman modes is symmetric in positive and negative magnetic fields, while it starts to become asymmetric below $\sim 150$ K. The strong asymmetry and appearance of the $S$ peak at temperatures less than $T_{\textit{N}}$ is clearly observed.

\begin{figure}[ht]
    \centering
    \includegraphics[width=0.9\linewidth]{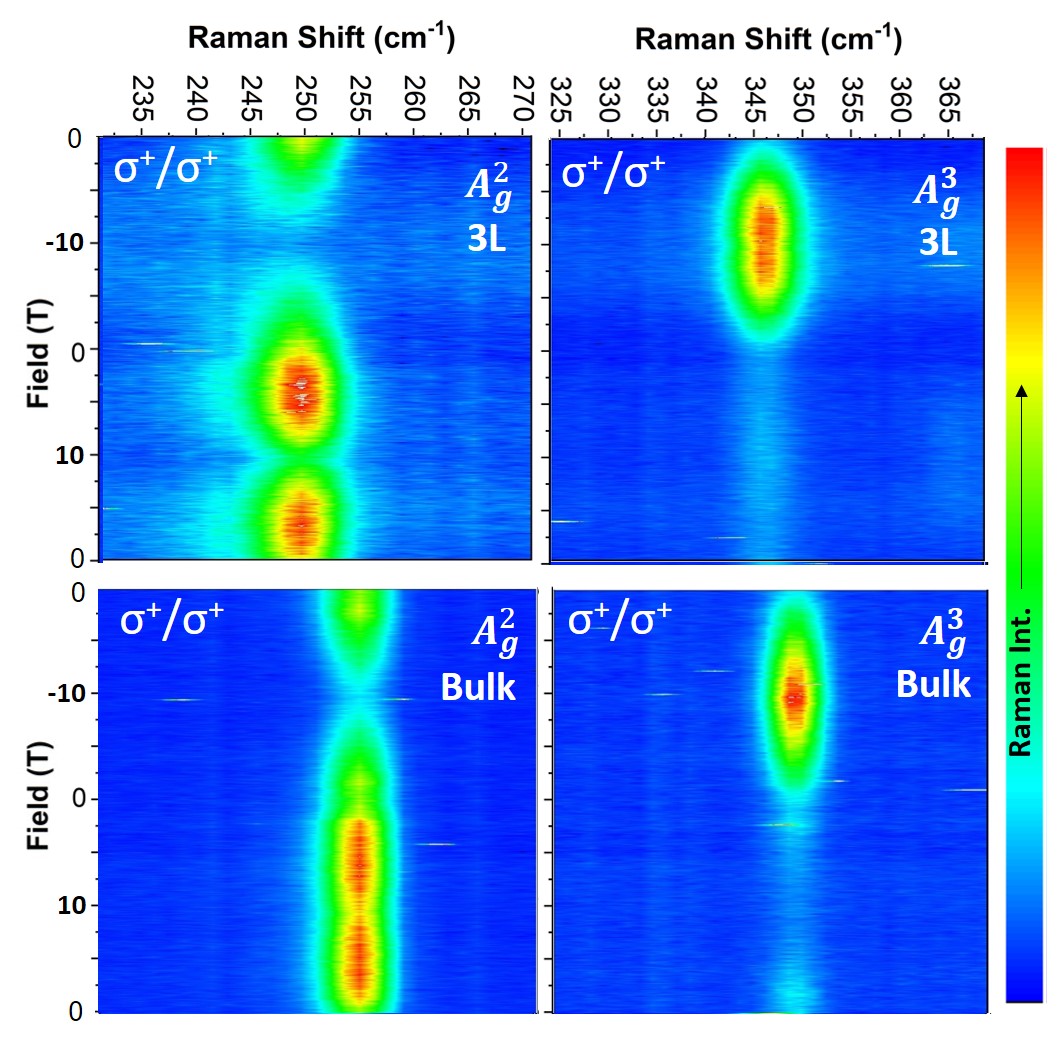}
    \caption{Contour maps of the Raman intensity vs. magnetic field and Raman shift frequency for $A_g^2$ and $A_g^3$ phonon modes of 3L and bulk CrSBr in $\sigma^{+}$/$\sigma^{+}$ parallel detection mode.}
\label{fig:FS5}
\end{figure}

Furthermore, 3L and bulk CrSBr were measured at 4~K in the magnetic field with $\sigma^{+}$/$\sigma^{+}$ parallel detection mode to investigate the layer thickness effect. The results of these measurements again show the asymmetric behavior of both layered and bulk CrSBr with respect to the magnetic-field scan, as depicted in Fig.~\ref{fig:FS5}. Here, we observe redshift of the $A_g^2$ and $A_g^3$ phonon frequencies with a decreasing number of layers but the characteristic peak $S$ is absent in the bulk CrSBr.     

\section{Summary and conclusions\label{summary}}

We have carried out systematic investigations of the magneto-Raman effect in the two-dimensional layered van der Waals CrSBr antiferromagnet by performing measurements as a function of temperature and applied magnetic field. Various forms of CrSBr system were examined, from bulk material to monolayer. The behavior of the Raman spectra, in particular those showing Raman-active modes of $A_g$ symmetry, has been studied upon different circular polarizations of the laser beam. Our experimental research is supported by \textit{ab initio} simulations of the Raman spectra which were undertaken to explore the effect of spin arrangement on the vibrational dynamics of the bulk and monolayer CrSBr. Both experiments and simulations reveal substantial effect of magnetic exchange interactions on the intensities and widths of the Raman-active phonon modes for several layers as well as bulk samples. Experimental magneto-Raman investigations unambiguously indicate that above T$_N$ a substantial decrease in the Raman shifts of the $A_g$ modes is accompanied by their significant broadening. This indicates that spin-spin and spin-phonon interactions occur below the N\'{e}el temperature and become stronger in the CrSBr monolayer. Spin-phonon coupling is responsible for the appearance of additional phonon peaks emerging below the N\'{e}el temperature in the Raman spectra only for 1-6 layers of CrSBr, but they are absent in its bulk form.  \textit{Ab initio} simulated Raman spectra when compared to those determined in experiments suggest that the spin magnetic moments are most likely aligned along the easy magnetization axis in the CrSBr antiferromagnet.  

\begin{acknowledgments}
The projects e-INFRA CZ (ID:90254) and LUABA22048 supported by the Ministry of Education, Youth and Sports of the Czech Republic and the project No. 25-14529L of the Czech Science Foundation are acknowledged. VV, JV, and DL acknowledge support from the QM4ST project, no. CZ$.02.01.01/00/22\_008/0004572$, funded by MEYS and co-funded by the EU. The authors acknowledge Z. Sofer for providing CrSBr crystals and MGML (Czech Research Infrastructures funded by the MEYS, project no. LM2023065) for providing facilities to perform the magneto-Raman experiments. 
\end{acknowledgments}

\bibliographystyle{apsrev4-2}
\bibliography{references.bib}

\begin{thebibliography}{54}%
\makeatletter
\providecommand \@ifxundefined [1]{%
 \@ifx{#1\undefined}
}%
\providecommand \@ifnum [1]{%
 \ifnum #1\expandafter \@firstoftwo
 \else \expandafter \@secondoftwo
 \fi
}%
\providecommand \@ifx [1]{%
 \ifx #1\expandafter \@firstoftwo
 \else \expandafter \@secondoftwo
 \fi
}%
\providecommand \natexlab [1]{#1}%
\providecommand \enquote  [1]{``#1''}%
\providecommand \bibnamefont  [1]{#1}%
\providecommand \bibfnamefont [1]{#1}%
\providecommand \citenamefont [1]{#1}%
\providecommand \href@noop [0]{\@secondoftwo}%
\providecommand \href [0]{\begingroup \@sanitize@url \@href}%
\providecommand \@href[1]{\@@startlink{#1}\@@href}%
\providecommand \@@href[1]{\endgroup#1\@@endlink}%
\providecommand \@sanitize@url [0]{\catcode `\\12\catcode `\$12\catcode
  `\&12\catcode `\#12\catcode `\^12\catcode `\_12\catcode `\%12\relax}%
\providecommand \@@startlink[1]{}%
\providecommand \@@endlink[0]{}%
\providecommand \url  [0]{\begingroup\@sanitize@url \@url }%
\providecommand \@url [1]{\endgroup\@href {#1}{\urlprefix }}%
\providecommand \urlprefix  [0]{URL }%
\providecommand \Eprint [0]{\href }%
\providecommand \doibase [0]{https://doi.org/}%
\providecommand \selectlanguage [0]{\@gobble}%
\providecommand \bibinfo  [0]{\@secondoftwo}%
\providecommand \bibfield  [0]{\@secondoftwo}%
\providecommand \translation [1]{[#1]}%
\providecommand \BibitemOpen [0]{}%
\providecommand \bibitemStop [0]{}%
\providecommand \bibitemNoStop [0]{.\EOS\space}%
\providecommand \EOS [0]{\spacefactor3000\relax}%
\providecommand \BibitemShut  [1]{\csname bibitem#1\endcsname}%
\let\auto@bib@innerbib\@empty
\bibitem [{\citenamefont {Burch}\ \emph {et~al.}(2018)\citenamefont {Burch},
  \citenamefont {Mandrus},\ and\ \citenamefont {Park}}]{Burch2018}%
  \BibitemOpen
  \bibfield  {author} {\bibinfo {author} {\bibfnamefont {K.~S.}\ \bibnamefont
  {Burch}}, \bibinfo {author} {\bibfnamefont {D.}~\bibnamefont {Mandrus}},\
  and\ \bibinfo {author} {\bibfnamefont {J.-G.}\ \bibnamefont {Park}},\
  }\href@noop {} {\bibfield  {journal} {\bibinfo  {journal} {Nature}\ }\textbf
  {\bibinfo {volume} {563}} (\bibinfo {year} {2018})}\BibitemShut {NoStop}%
\bibitem [{\citenamefont {Huang}\ \emph {et~al.}(2017)\citenamefont {Huang},
  \citenamefont {Clark}, \citenamefont {Navarro-Moratalla}, \citenamefont
  {Klein}, \citenamefont {Cheng}, \citenamefont {Seyler}, \citenamefont
  {Zhong}, \citenamefont {Schmidgall}, \citenamefont {McGuire}, \citenamefont
  {Cobden}, \citenamefont {Yao}, \citenamefont {Xiao}, \citenamefont
  {Jarillo-Herrero},\ and\ \citenamefont {Xu}}]{Huang2017}%
  \BibitemOpen
  \bibfield  {author} {\bibinfo {author} {\bibfnamefont {B.}~\bibnamefont
  {Huang}}, \bibinfo {author} {\bibfnamefont {G.}~\bibnamefont {Clark}},
  \bibinfo {author} {\bibfnamefont {E.}~\bibnamefont {Navarro-Moratalla}},
  \bibinfo {author} {\bibfnamefont {D.~R.}\ \bibnamefont {Klein}}, \bibinfo
  {author} {\bibfnamefont {R.}~\bibnamefont {Cheng}}, \bibinfo {author}
  {\bibfnamefont {K.~L.}\ \bibnamefont {Seyler}}, \bibinfo {author}
  {\bibfnamefont {D.}~\bibnamefont {Zhong}}, \bibinfo {author} {\bibfnamefont
  {E.}~\bibnamefont {Schmidgall}}, \bibinfo {author} {\bibfnamefont {M.~A.}\
  \bibnamefont {McGuire}}, \bibinfo {author} {\bibfnamefont {D.~H.}\
  \bibnamefont {Cobden}}, \bibinfo {author} {\bibfnamefont {W.}~\bibnamefont
  {Yao}}, \bibinfo {author} {\bibfnamefont {D.}~\bibnamefont {Xiao}}, \bibinfo
  {author} {\bibfnamefont {P.}~\bibnamefont {Jarillo-Herrero}},\ and\ \bibinfo
  {author} {\bibfnamefont {X.}~\bibnamefont {Xu}},\ }\href@noop {} {\bibfield
  {journal} {\bibinfo  {journal} {Nature}\ }\textbf {\bibinfo {volume} {546}}
  (\bibinfo {year} {2017})}\BibitemShut {NoStop}%
\bibitem [{\citenamefont {Gong}\ \emph {et~al.}(2017)\citenamefont {Gong},
  \citenamefont {Li}, \citenamefont {Li}, \citenamefont {Ji}, \citenamefont
  {Stern}, \citenamefont {Xia}, \citenamefont {Cao}, \citenamefont {Bao},
  \citenamefont {Wang}, \citenamefont {Wang}, \citenamefont {Qiu},
  \citenamefont {Cava}, \citenamefont {Louie}, \citenamefont {Xia},\ and\
  \citenamefont {Zhang}}]{Gong2017}%
  \BibitemOpen
  \bibfield  {author} {\bibinfo {author} {\bibfnamefont {C.}~\bibnamefont
  {Gong}}, \bibinfo {author} {\bibfnamefont {L.}~\bibnamefont {Li}}, \bibinfo
  {author} {\bibfnamefont {Z.}~\bibnamefont {Li}}, \bibinfo {author}
  {\bibfnamefont {H.}~\bibnamefont {Ji}}, \bibinfo {author} {\bibfnamefont
  {A.}~\bibnamefont {Stern}}, \bibinfo {author} {\bibfnamefont
  {Y.}~\bibnamefont {Xia}}, \bibinfo {author} {\bibfnamefont {T.}~\bibnamefont
  {Cao}}, \bibinfo {author} {\bibfnamefont {W.}~\bibnamefont {Bao}}, \bibinfo
  {author} {\bibfnamefont {C.}~\bibnamefont {Wang}}, \bibinfo {author}
  {\bibfnamefont {Y.}~\bibnamefont {Wang}}, \bibinfo {author} {\bibfnamefont
  {Z.~Q.}\ \bibnamefont {Qiu}}, \bibinfo {author} {\bibfnamefont {R.~J.}\
  \bibnamefont {Cava}}, \bibinfo {author} {\bibfnamefont {S.~G.}\ \bibnamefont
  {Louie}}, \bibinfo {author} {\bibfnamefont {J.}~\bibnamefont {Xia}},\ and\
  \bibinfo {author} {\bibfnamefont {X.}~\bibnamefont {Zhang}},\ }\href@noop {}
  {\bibfield  {journal} {\bibinfo  {journal} {Nature}\ }\textbf {\bibinfo
  {volume} {546}} (\bibinfo {year} {2017})}\BibitemShut {NoStop}%
\bibitem [{\citenamefont {Tu}\ \emph {et~al.}(2022)\citenamefont {Tu},
  \citenamefont {Liu3}, \citenamefont {Hou}, \citenamefont {Shi}, \citenamefont
  {Jia}, \citenamefont {Su}, \citenamefont {Zhang}, \citenamefont {Zhang},\
  and\ \citenamefont {Wang}}]{Tu2022}%
  \BibitemOpen
  \bibfield  {author} {\bibinfo {author} {\bibfnamefont {Y.}~\bibnamefont
  {Tu}}, \bibinfo {author} {\bibfnamefont {Q.}~\bibnamefont {Liu3}}, \bibinfo
  {author} {\bibfnamefont {L.}~\bibnamefont {Hou}}, \bibinfo {author}
  {\bibfnamefont {P.}~\bibnamefont {Shi}}, \bibinfo {author} {\bibfnamefont
  {C.}~\bibnamefont {Jia}}, \bibinfo {author} {\bibfnamefont {J.}~\bibnamefont
  {Su}}, \bibinfo {author} {\bibfnamefont {J.}~\bibnamefont {Zhang}}, \bibinfo
  {author} {\bibfnamefont {X.}~\bibnamefont {Zhang}},\ and\ \bibinfo {author}
  {\bibfnamefont {B.}~\bibnamefont {Wang}},\ }\href@noop {} {\bibfield
  {journal} {\bibinfo  {journal} {Front. Phys.}\ }\textbf {\bibinfo {volume}
  {10}} (\bibinfo {year} {2022})}\BibitemShut {NoStop}%
\bibitem [{\citenamefont {Sivadas}\ \emph {et~al.}(2018)\citenamefont
  {Sivadas}, \citenamefont {Okamoto}, \citenamefont {Xu}, \citenamefont
  {Fennie},\ and\ \citenamefont {Xiao}}]{Sivadas2018}%
  \BibitemOpen
  \bibfield  {author} {\bibinfo {author} {\bibfnamefont {N.}~\bibnamefont
  {Sivadas}}, \bibinfo {author} {\bibfnamefont {S.}~\bibnamefont {Okamoto}},
  \bibinfo {author} {\bibfnamefont {X.}~\bibnamefont {Xu}}, \bibinfo {author}
  {\bibfnamefont {C.~J.}\ \bibnamefont {Fennie}},\ and\ \bibinfo {author}
  {\bibfnamefont {D.}~\bibnamefont {Xiao}},\ }\href@noop {} {\bibfield
  {journal} {\bibinfo  {journal} {Nano Lett.}\ }\textbf {\bibinfo {volume}
  {18}} (\bibinfo {year} {2018})}\BibitemShut {NoStop}%
\bibitem [{\citenamefont {Huang}\ \emph {et~al.}(2018)\citenamefont {Huang},
  \citenamefont {Feng}, \citenamefont {Wu}, \citenamefont {Ahmed},
  \citenamefont {Huang}, \citenamefont {Xiang}, \citenamefont {Deng},\ and\
  \citenamefont {Kan}}]{Huang2018}%
  \BibitemOpen
  \bibfield  {author} {\bibinfo {author} {\bibfnamefont {C.}~\bibnamefont
  {Huang}}, \bibinfo {author} {\bibfnamefont {J.}~\bibnamefont {Feng}},
  \bibinfo {author} {\bibfnamefont {F.}~\bibnamefont {Wu}}, \bibinfo {author}
  {\bibfnamefont {D.}~\bibnamefont {Ahmed}}, \bibinfo {author} {\bibfnamefont
  {B.}~\bibnamefont {Huang}}, \bibinfo {author} {\bibfnamefont
  {H.}~\bibnamefont {Xiang}}, \bibinfo {author} {\bibfnamefont
  {K.}~\bibnamefont {Deng}},\ and\ \bibinfo {author} {\bibfnamefont
  {E.}~\bibnamefont {Kan}},\ }\href@noop {} {\bibfield  {journal} {\bibinfo
  {journal} {J. Am. Chem. Soc.}\ }\textbf {\bibinfo {volume} {10}} (\bibinfo
  {year} {2018})}\BibitemShut {NoStop}%
\bibitem [{\citenamefont {Wang}\ \emph {et~al.}(2012)\citenamefont {Wang},
  \citenamefont {Kalantar-Zadeh}, \citenamefont {Kis}, \citenamefont
  {Coleman},\ and\ \citenamefont {Strano}}]{Wang2012}%
  \BibitemOpen
  \bibfield  {author} {\bibinfo {author} {\bibfnamefont {Q.~H.}\ \bibnamefont
  {Wang}}, \bibinfo {author} {\bibfnamefont {K.}~\bibnamefont
  {Kalantar-Zadeh}}, \bibinfo {author} {\bibfnamefont {A.}~\bibnamefont {Kis}},
  \bibinfo {author} {\bibfnamefont {J.~N.}\ \bibnamefont {Coleman}},\ and\
  \bibinfo {author} {\bibfnamefont {M.~S.}\ \bibnamefont {Strano}},\
  }\href@noop {} {\bibfield  {journal} {\bibinfo  {journal} {Nat.
  Nanotechnol.}\ }\textbf {\bibinfo {volume} {7}} (\bibinfo {year}
  {2012})}\BibitemShut {NoStop}%
\bibitem [{\citenamefont {Lin}\ \emph {et~al.}(2018)\citenamefont {Lin},
  \citenamefont {Liu}, \citenamefont {Halim}, \citenamefont {Ding},
  \citenamefont {Liu}, \citenamefont {Wang}, \citenamefont {Jia}, \citenamefont
  {Chen}, \citenamefont {Duan}, \citenamefont {Wang}, \citenamefont {Song},
  \citenamefont {Li}, \citenamefont {Wan}, \citenamefont {Huang},\ and\
  \citenamefont {Duan}}]{Lin2018}%
  \BibitemOpen
  \bibfield  {author} {\bibinfo {author} {\bibfnamefont {Z.}~\bibnamefont
  {Lin}}, \bibinfo {author} {\bibfnamefont {Y.}~\bibnamefont {Liu}}, \bibinfo
  {author} {\bibfnamefont {U.}~\bibnamefont {Halim}}, \bibinfo {author}
  {\bibfnamefont {M.}~\bibnamefont {Ding}}, \bibinfo {author} {\bibfnamefont
  {Y.}~\bibnamefont {Liu}}, \bibinfo {author} {\bibfnamefont {Y.}~\bibnamefont
  {Wang}}, \bibinfo {author} {\bibfnamefont {C.}~\bibnamefont {Jia}}, \bibinfo
  {author} {\bibfnamefont {P.}~\bibnamefont {Chen}}, \bibinfo {author}
  {\bibfnamefont {X.}~\bibnamefont {Duan}}, \bibinfo {author} {\bibfnamefont
  {C.}~\bibnamefont {Wang}}, \bibinfo {author} {\bibfnamefont {F.}~\bibnamefont
  {Song}}, \bibinfo {author} {\bibfnamefont {M.}~\bibnamefont {Li}}, \bibinfo
  {author} {\bibfnamefont {C.}~\bibnamefont {Wan}}, \bibinfo {author}
  {\bibfnamefont {Y.}~\bibnamefont {Huang}},\ and\ \bibinfo {author}
  {\bibfnamefont {X.}~\bibnamefont {Duan}},\ }\href@noop {} {\bibfield
  {journal} {\bibinfo  {journal} {Nature}\ }\textbf {\bibinfo {volume} {562}}
  (\bibinfo {year} {2018})}\BibitemShut {NoStop}%
\bibitem [{\citenamefont {Shim}\ \emph {et~al.}(2017)\citenamefont {Shim},
  \citenamefont {Park}, \citenamefont {Kang}, \citenamefont {Kim},
  \citenamefont {Jo}, \citenamefont {Park},\ and\ \citenamefont
  {Park}}]{Shim2017}%
  \BibitemOpen
  \bibfield  {author} {\bibinfo {author} {\bibfnamefont {J.}~\bibnamefont
  {Shim}}, \bibinfo {author} {\bibfnamefont {H.-Y.}\ \bibnamefont {Park}},
  \bibinfo {author} {\bibfnamefont {D.-H.}\ \bibnamefont {Kang}}, \bibinfo
  {author} {\bibfnamefont {J.-O.}\ \bibnamefont {Kim}}, \bibinfo {author}
  {\bibfnamefont {S.-H.}\ \bibnamefont {Jo}}, \bibinfo {author} {\bibfnamefont
  {Y.}~\bibnamefont {Park}},\ and\ \bibinfo {author} {\bibfnamefont {J.-H.}\
  \bibnamefont {Park}},\ }\href@noop {} {\bibfield  {journal} {\bibinfo
  {journal} {Ad. Electron. Mater.}\ }\textbf {\bibinfo {volume} {3}} (\bibinfo
  {year} {2017})}\BibitemShut {NoStop}%
\bibitem [{\citenamefont {Liu}\ \emph {et~al.}(2017)\citenamefont {Liu},
  \citenamefont {Abbas},\ and\ \citenamefont {Zhou}}]{Liu2017}%
  \BibitemOpen
  \bibfield  {author} {\bibinfo {author} {\bibfnamefont {B.}~\bibnamefont
  {Liu}}, \bibinfo {author} {\bibfnamefont {A.}~\bibnamefont {Abbas}},\ and\
  \bibinfo {author} {\bibfnamefont {C.}~\bibnamefont {Zhou}},\ }\href@noop {}
  {\bibfield  {journal} {\bibinfo  {journal} {Ad. Electron. Mater.}\ }\textbf
  {\bibinfo {volume} {3}} (\bibinfo {year} {2017})}\BibitemShut {NoStop}%
\bibitem [{\citenamefont {Tan}\ \emph {et~al.}(2020)\citenamefont {Tan},
  \citenamefont {Jiang}, \citenamefont {Wang}, \citenamefont {Yao},\ and\
  \citenamefont {Zhang}}]{Tan2020}%
  \BibitemOpen
  \bibfield  {author} {\bibinfo {author} {\bibfnamefont {T.}~\bibnamefont
  {Tan}}, \bibinfo {author} {\bibfnamefont {X.}~\bibnamefont {Jiang}}, \bibinfo
  {author} {\bibfnamefont {C.}~\bibnamefont {Wang}}, \bibinfo {author}
  {\bibfnamefont {B.}~\bibnamefont {Yao}},\ and\ \bibinfo {author}
  {\bibfnamefont {H.}~\bibnamefont {Zhang}},\ }\href@noop {} {\bibfield
  {journal} {\bibinfo  {journal} {Adv. Sci.}\ }\textbf {\bibinfo {volume} {7}}
  (\bibinfo {year} {2020})}\BibitemShut {NoStop}%
\bibitem [{\citenamefont {Guo}\ \emph {et~al.}(2018)\citenamefont {Guo},
  \citenamefont {Zhang}, \citenamefont {Yuan}, \citenamefont {Wang},\ and\
  \citenamefont {Wang}}]{guo18}%
  \BibitemOpen
  \bibfield  {author} {\bibinfo {author} {\bibfnamefont {Y.}~\bibnamefont
  {Guo}}, \bibinfo {author} {\bibfnamefont {Y.}~\bibnamefont {Zhang}}, \bibinfo
  {author} {\bibfnamefont {S.}~\bibnamefont {Yuan}}, \bibinfo {author}
  {\bibfnamefont {B.}~\bibnamefont {Wang}},\ and\ \bibinfo {author}
  {\bibfnamefont {J.}~\bibnamefont {Wang}},\ }\href@noop {} {\bibfield
  {journal} {\bibinfo  {journal} {Nanoscale}\ }\textbf {\bibinfo {volume}
  {10}},\ \bibinfo {pages} {18036} (\bibinfo {year} {2018})}\BibitemShut
  {NoStop}%
\bibitem [{\citenamefont {Telford}\ \emph {et~al.}(2020)\citenamefont
  {Telford}, \citenamefont {Dismukes}, \citenamefont {Lee}, \citenamefont
  {Cheng}, \citenamefont {nad A.~K.~Bartholomew}, \citenamefont {Chen},
  \citenamefont {Xu}, \citenamefont {Pasupathy}, \citenamefont {Zhu},
  \citenamefont {Dean},\ and\ \citenamefont {Roy}}]{Telford2020}%
  \BibitemOpen
  \bibfield  {author} {\bibinfo {author} {\bibfnamefont {E.~J.}\ \bibnamefont
  {Telford}}, \bibinfo {author} {\bibfnamefont {A.~H.}\ \bibnamefont
  {Dismukes}}, \bibinfo {author} {\bibfnamefont {K.}~\bibnamefont {Lee}},
  \bibinfo {author} {\bibfnamefont {M.}~\bibnamefont {Cheng}}, \bibinfo
  {author} {\bibfnamefont {A.~W.}\ \bibnamefont {nad A.~K.~Bartholomew}},
  \bibinfo {author} {\bibfnamefont {Y.-S.}\ \bibnamefont {Chen}}, \bibinfo
  {author} {\bibfnamefont {X.}~\bibnamefont {Xu}}, \bibinfo {author}
  {\bibfnamefont {A.~N.}\ \bibnamefont {Pasupathy}}, \bibinfo {author}
  {\bibfnamefont {X.}~\bibnamefont {Zhu}}, \bibinfo {author} {\bibfnamefont
  {C.~R.}\ \bibnamefont {Dean}},\ and\ \bibinfo {author} {\bibfnamefont
  {X.}~\bibnamefont {Roy}},\ }\href@noop {} {\bibfield  {journal} {\bibinfo
  {journal} {Adv. Matter.}\ }\textbf {\bibinfo {volume} {32}},\ \bibinfo
  {pages} {2003240} (\bibinfo {year} {2020})}\BibitemShut {NoStop}%
\bibitem [{\citenamefont {Lee}\ \emph {et~al.}(2021)\citenamefont {Lee},
  \citenamefont {Dismukes}, \citenamefont {Telford}, \citenamefont {Wiscons},
  \citenamefont {Wang}, \citenamefont {Xu}, \citenamefont {Nuckolls},
  \citenamefont {Dean}, \citenamefont {Roy},\ and\ \citenamefont
  {Zhu}}]{Lee2021}%
  \BibitemOpen
  \bibfield  {author} {\bibinfo {author} {\bibfnamefont {K.}~\bibnamefont
  {Lee}}, \bibinfo {author} {\bibfnamefont {A.~H.}\ \bibnamefont {Dismukes}},
  \bibinfo {author} {\bibfnamefont {E.~J.}\ \bibnamefont {Telford}}, \bibinfo
  {author} {\bibfnamefont {R.~A.}\ \bibnamefont {Wiscons}}, \bibinfo {author}
  {\bibfnamefont {J.}~\bibnamefont {Wang}}, \bibinfo {author} {\bibfnamefont
  {X.}~\bibnamefont {Xu}}, \bibinfo {author} {\bibfnamefont {C.}~\bibnamefont
  {Nuckolls}}, \bibinfo {author} {\bibfnamefont {C.~R.}\ \bibnamefont {Dean}},
  \bibinfo {author} {\bibfnamefont {X.}~\bibnamefont {Roy}},\ and\ \bibinfo
  {author} {\bibfnamefont {X.}~\bibnamefont {Zhu}},\ }\href@noop {} {\bibfield
  {journal} {\bibinfo  {journal} {Nano Lett.}\ }\textbf {\bibinfo {volume}
  {21}} (\bibinfo {year} {2021})}\BibitemShut {NoStop}%
\bibitem [{\citenamefont {Yang}\ \emph {et~al.}(2021)\citenamefont {Yang},
  \citenamefont {Wang}, \citenamefont {Liu}, \citenamefont {Lu},\ and\
  \citenamefont {Wu}}]{yang21}%
  \BibitemOpen
  \bibfield  {author} {\bibinfo {author} {\bibfnamefont {K.}~\bibnamefont
  {Yang}}, \bibinfo {author} {\bibfnamefont {G.}~\bibnamefont {Wang}}, \bibinfo
  {author} {\bibfnamefont {L.}~\bibnamefont {Liu}}, \bibinfo {author}
  {\bibfnamefont {D.}~\bibnamefont {Lu}},\ and\ \bibinfo {author}
  {\bibfnamefont {H.}~\bibnamefont {Wu}},\ }\href@noop {} {\bibfield  {journal}
  {\bibinfo  {journal} {Phys. Rev. B}\ }\textbf {\bibinfo {volume} {104}},\
  \bibinfo {pages} {144416} (\bibinfo {year} {2021})}\BibitemShut {NoStop}%
\bibitem [{\citenamefont {Wilson}\ \emph {et~al.}(2021)\citenamefont {Wilson},
  \citenamefont {Lee}, \citenamefont {Cenker}, \citenamefont {Xie},
  \citenamefont {Dismukes}, \citenamefont {Telford}, \citenamefont {Fonseca},
  \citenamefont {Sivakumar}, \citenamefont {Dean}, \citenamefont {Cao},
  \citenamefont {Roy}, \citenamefont {Xu}, ,\ and\ \citenamefont
  {Zhu}}]{Wilson21}%
  \BibitemOpen
  \bibfield  {author} {\bibinfo {author} {\bibfnamefont {N.~P.}\ \bibnamefont
  {Wilson}}, \bibinfo {author} {\bibfnamefont {K.}~\bibnamefont {Lee}},
  \bibinfo {author} {\bibfnamefont {J.}~\bibnamefont {Cenker}}, \bibinfo
  {author} {\bibfnamefont {K.}~\bibnamefont {Xie}}, \bibinfo {author}
  {\bibfnamefont {A.~H.}\ \bibnamefont {Dismukes}}, \bibinfo {author}
  {\bibfnamefont {E.~J.}\ \bibnamefont {Telford}}, \bibinfo {author}
  {\bibfnamefont {J.}~\bibnamefont {Fonseca}}, \bibinfo {author} {\bibfnamefont
  {S.}~\bibnamefont {Sivakumar}}, \bibinfo {author} {\bibfnamefont
  {C.}~\bibnamefont {Dean}}, \bibinfo {author} {\bibfnamefont {T.}~\bibnamefont
  {Cao}}, \bibinfo {author} {\bibfnamefont {X.}~\bibnamefont {Roy}}, \bibinfo
  {author} {\bibfnamefont {X.}~\bibnamefont {Xu}}, ,\ and\ \bibinfo {author}
  {\bibfnamefont {X.}~\bibnamefont {Zhu}},\ }\href@noop {} {\bibfield
  {journal} {\bibinfo  {journal} {Nature Mater.}\ }\textbf {\bibinfo {volume}
  {20}},\ \bibinfo {pages} {1657} (\bibinfo {year} {2021})}\BibitemShut
  {NoStop}%
\bibitem [{\citenamefont {Xu}\ \emph {et~al.}(2022)\citenamefont {Xu},
  \citenamefont {Wang}, \citenamefont {Chang}, \citenamefont {Chen},
  \citenamefont {Guan},\ and\ \citenamefont {Tao}}]{Xu2022}%
  \BibitemOpen
  \bibfield  {author} {\bibinfo {author} {\bibfnamefont {X.}~\bibnamefont
  {Xu}}, \bibinfo {author} {\bibfnamefont {X.}~\bibnamefont {Wang}}, \bibinfo
  {author} {\bibfnamefont {P.}~\bibnamefont {Chang}}, \bibinfo {author}
  {\bibfnamefont {X.}~\bibnamefont {Chen}}, \bibinfo {author} {\bibfnamefont
  {L.}~\bibnamefont {Guan}},\ and\ \bibinfo {author} {\bibfnamefont
  {J.}~\bibnamefont {Tao}},\ }\href@noop {} {\bibfield  {journal} {\bibinfo
  {journal} {J. Phys. Chem. C}\ }\textbf {\bibinfo {volume} {126}},\ \bibinfo
  {pages} {10574} (\bibinfo {year} {2022})}\BibitemShut {NoStop}%
\bibitem [{\citenamefont {Klein}\ \emph {et~al.}(2022)\citenamefont {Klein},
  \citenamefont {Pham}, \citenamefont {Thomsen}, \citenamefont {Curtis},
  \citenamefont {Denneulin}, \citenamefont {Lorke}, \citenamefont {Florian},
  \citenamefont {Steinhoff}, \citenamefont {Wiscons}, \citenamefont {Luxa},
  \citenamefont {Sofer}, \citenamefont {Jahnke}, \citenamefont {Narang},\ and\
  \citenamefont {Ross}}]{Klein2022}%
  \BibitemOpen
  \bibfield  {author} {\bibinfo {author} {\bibfnamefont {J.}~\bibnamefont
  {Klein}}, \bibinfo {author} {\bibfnamefont {T.}~\bibnamefont {Pham}},
  \bibinfo {author} {\bibfnamefont {J.~D.}\ \bibnamefont {Thomsen}}, \bibinfo
  {author} {\bibfnamefont {J.~B.}\ \bibnamefont {Curtis}}, \bibinfo {author}
  {\bibfnamefont {T.}~\bibnamefont {Denneulin}}, \bibinfo {author}
  {\bibfnamefont {M.}~\bibnamefont {Lorke}}, \bibinfo {author} {\bibfnamefont
  {M.}~\bibnamefont {Florian}}, \bibinfo {author} {\bibfnamefont
  {A.}~\bibnamefont {Steinhoff}}, \bibinfo {author} {\bibfnamefont {R.~A.}\
  \bibnamefont {Wiscons}}, \bibinfo {author} {\bibfnamefont {J.}~\bibnamefont
  {Luxa}}, \bibinfo {author} {\bibfnamefont {Z.}~\bibnamefont {Sofer}},
  \bibinfo {author} {\bibfnamefont {F.}~\bibnamefont {Jahnke}}, \bibinfo
  {author} {\bibfnamefont {P.}~\bibnamefont {Narang}},\ and\ \bibinfo {author}
  {\bibfnamefont {F.~M.}\ \bibnamefont {Ross}},\ }\href@noop {} {\bibfield
  {journal} {\bibinfo  {journal} {Nature Commun.}\ }\textbf {\bibinfo {volume}
  {13}},\ \bibinfo {pages} {5420} (\bibinfo {year} {2022})}\BibitemShut
  {NoStop}%
\bibitem [{\citenamefont {López-Paz}\ \emph {et~al.}(2022)\citenamefont
  {López-Paz}, \citenamefont {Guguchia}, \citenamefont {Pomjakushin},
  \citenamefont {Witteveen}, \citenamefont {Cervellino}, \citenamefont
  {H.~Luetkens}, \citenamefont {Morpurgo},\ and\ \citenamefont {von
  Rohr}}]{lopez22}%
  \BibitemOpen
  \bibfield  {author} {\bibinfo {author} {\bibfnamefont {S.~A.}\ \bibnamefont
  {López-Paz}}, \bibinfo {author} {\bibfnamefont {Z.}~\bibnamefont
  {Guguchia}}, \bibinfo {author} {\bibfnamefont {V.~Y.}\ \bibnamefont
  {Pomjakushin}}, \bibinfo {author} {\bibfnamefont {C.}~\bibnamefont
  {Witteveen}}, \bibinfo {author} {\bibfnamefont {A.}~\bibnamefont
  {Cervellino}}, \bibinfo {author} {\bibfnamefont {N.~C.}\ \bibnamefont
  {H.~Luetkens}}, \bibinfo {author} {\bibfnamefont {A.~F.}\ \bibnamefont
  {Morpurgo}},\ and\ \bibinfo {author} {\bibfnamefont {F.~O.}\ \bibnamefont
  {von Rohr}},\ }\href@noop {} {\bibfield  {journal} {\bibinfo  {journal} {Nat.
  Commun.}\ }\textbf {\bibinfo {volume} {13}},\ \bibinfo {pages} {4745}
  (\bibinfo {year} {2022})}\BibitemShut {NoStop}%
\bibitem [{\citenamefont {Klein}\ \emph
  {et~al.}(2023{\natexlab{a}})\citenamefont {Klein}, \citenamefont {Song},
  \citenamefont {Pingault}, \citenamefont {Dirnberger}, \citenamefont {Chi},
  \citenamefont {Curtis}, \citenamefont {Dana}, \citenamefont {Bushati},
  \citenamefont {Quan}, \citenamefont {Dekanovsky}, \citenamefont {Sofer},
  \citenamefont {Alù}, \citenamefont {Menon}, \citenamefont {Moodera},
  \citenamefont {Lončar}, \citenamefont {Narang},\ and\ \citenamefont
  {Ross}}]{Klein23a}%
  \BibitemOpen
  \bibfield  {author} {\bibinfo {author} {\bibfnamefont {J.}~\bibnamefont
  {Klein}}, \bibinfo {author} {\bibfnamefont {Z.}~\bibnamefont {Song}},
  \bibinfo {author} {\bibfnamefont {B.}~\bibnamefont {Pingault}}, \bibinfo
  {author} {\bibfnamefont {F.}~\bibnamefont {Dirnberger}}, \bibinfo {author}
  {\bibfnamefont {H.}~\bibnamefont {Chi}}, \bibinfo {author} {\bibfnamefont
  {J.~B.}\ \bibnamefont {Curtis}}, \bibinfo {author} {\bibfnamefont
  {R.}~\bibnamefont {Dana}}, \bibinfo {author} {\bibfnamefont {R.}~\bibnamefont
  {Bushati}}, \bibinfo {author} {\bibfnamefont {J.}~\bibnamefont {Quan}},
  \bibinfo {author} {\bibfnamefont {L.}~\bibnamefont {Dekanovsky}}, \bibinfo
  {author} {\bibfnamefont {Z.}~\bibnamefont {Sofer}}, \bibinfo {author}
  {\bibfnamefont {A.}~\bibnamefont {Alù}}, \bibinfo {author} {\bibfnamefont
  {V.~M.}\ \bibnamefont {Menon}}, \bibinfo {author} {\bibfnamefont {J.~S.}\
  \bibnamefont {Moodera}}, \bibinfo {author} {\bibfnamefont {M.}~\bibnamefont
  {Lončar}}, \bibinfo {author} {\bibfnamefont {P.}~\bibnamefont {Narang}},\
  and\ \bibinfo {author} {\bibfnamefont {F.~M.}\ \bibnamefont {Ross}},\
  }\href@noop {} {\bibfield  {journal} {\bibinfo  {journal} {ACS Nano}\
  }\textbf {\bibinfo {volume} {17}},\ \bibinfo {pages} {288} (\bibinfo {year}
  {2023}{\natexlab{a}})}\BibitemShut {NoStop}%
\bibitem [{\citenamefont {Klein}\ \emph
  {et~al.}(2023{\natexlab{b}})\citenamefont {Klein}, \citenamefont {Pingault},
  \citenamefont {Florian}, \citenamefont {Heißenbüttel}, \citenamefont
  {Steinhoff}, \citenamefont {Song}, \citenamefont {Torres}, \citenamefont
  {Dirnberger}, \citenamefont {Curtis}, \citenamefont {Weile}, \citenamefont
  {Penn}, \citenamefont {Deilmann}, \citenamefont {Dana}, \citenamefont
  {Bushati}, \citenamefont {Quan}, \citenamefont {Luxa}, \citenamefont {Sofer},
  \citenamefont {Alù}, \citenamefont {Menon}, \citenamefont {Wurstbauer},
  \citenamefont {Rohlfing}, \citenamefont {Narang}, \citenamefont {Lončar},\
  and\ \citenamefont {Ross}}]{Klein23b}%
  \BibitemOpen
  \bibfield  {author} {\bibinfo {author} {\bibfnamefont {J.}~\bibnamefont
  {Klein}}, \bibinfo {author} {\bibfnamefont {B.}~\bibnamefont {Pingault}},
  \bibinfo {author} {\bibfnamefont {M.}~\bibnamefont {Florian}}, \bibinfo
  {author} {\bibfnamefont {M.-C.}\ \bibnamefont {Heißenbüttel}}, \bibinfo
  {author} {\bibfnamefont {A.}~\bibnamefont {Steinhoff}}, \bibinfo {author}
  {\bibfnamefont {Z.}~\bibnamefont {Song}}, \bibinfo {author} {\bibfnamefont
  {K.}~\bibnamefont {Torres}}, \bibinfo {author} {\bibfnamefont
  {F.}~\bibnamefont {Dirnberger}}, \bibinfo {author} {\bibfnamefont {J.~B.}\
  \bibnamefont {Curtis}}, \bibinfo {author} {\bibfnamefont {M.}~\bibnamefont
  {Weile}}, \bibinfo {author} {\bibfnamefont {A.}~\bibnamefont {Penn}},
  \bibinfo {author} {\bibfnamefont {T.}~\bibnamefont {Deilmann}}, \bibinfo
  {author} {\bibfnamefont {R.}~\bibnamefont {Dana}}, \bibinfo {author}
  {\bibfnamefont {R.}~\bibnamefont {Bushati}}, \bibinfo {author} {\bibfnamefont
  {J.}~\bibnamefont {Quan}}, \bibinfo {author} {\bibfnamefont {J.}~\bibnamefont
  {Luxa}}, \bibinfo {author} {\bibfnamefont {Z.}~\bibnamefont {Sofer}},
  \bibinfo {author} {\bibfnamefont {A.}~\bibnamefont {Alù}}, \bibinfo {author}
  {\bibfnamefont {V.~M.}\ \bibnamefont {Menon}}, \bibinfo {author}
  {\bibfnamefont {U.}~\bibnamefont {Wurstbauer}}, \bibinfo {author}
  {\bibfnamefont {M.}~\bibnamefont {Rohlfing}}, \bibinfo {author}
  {\bibfnamefont {P.}~\bibnamefont {Narang}}, \bibinfo {author} {\bibfnamefont
  {M.}~\bibnamefont {Lončar}},\ and\ \bibinfo {author} {\bibfnamefont {F.~M.}\
  \bibnamefont {Ross}},\ }\href@noop {} {\bibfield  {journal} {\bibinfo
  {journal} {ACS Nano}\ }\textbf {\bibinfo {volume} {17}},\ \bibinfo {pages}
  {5316} (\bibinfo {year} {2023}{\natexlab{b}})}\BibitemShut {NoStop}%
\bibitem [{\citenamefont {Pawbake}\ \emph {et~al.}(2023)\citenamefont
  {Pawbake}, \citenamefont {Pelini}, \citenamefont {Wilson}, \citenamefont
  {Mosina}, \citenamefont {Sofer}, \citenamefont {Heid},\ and\ \citenamefont
  {Faugeras}}]{pawbake23}%
  \BibitemOpen
  \bibfield  {author} {\bibinfo {author} {\bibfnamefont {A.}~\bibnamefont
  {Pawbake}}, \bibinfo {author} {\bibfnamefont {T.}~\bibnamefont {Pelini}},
  \bibinfo {author} {\bibfnamefont {N.~P.}\ \bibnamefont {Wilson}}, \bibinfo
  {author} {\bibfnamefont {K.}~\bibnamefont {Mosina}}, \bibinfo {author}
  {\bibfnamefont {Z.}~\bibnamefont {Sofer}}, \bibinfo {author} {\bibfnamefont
  {R.}~\bibnamefont {Heid}},\ and\ \bibinfo {author} {\bibfnamefont
  {C.}~\bibnamefont {Faugeras}},\ }\href@noop {} {\bibfield  {journal}
  {\bibinfo  {journal} {Phys. Rev. B}\ }\textbf {\bibinfo {volume} {107}},\
  \bibinfo {pages} {075421} (\bibinfo {year} {2023})}\BibitemShut {NoStop}%
\bibitem [{\citenamefont {Linhart}\ \emph {et~al.}(2023)\citenamefont
  {Linhart}, \citenamefont {Rybak}, \citenamefont {Birowska}, \citenamefont
  {Scharoch}, \citenamefont {Mosina}, \citenamefont {Mazanek}, \citenamefont
  {Kaczorowski}, \citenamefont {Sofer},\ and\ \citenamefont
  {Kudrawiec}}]{Linhart2023}%
  \BibitemOpen
  \bibfield  {author} {\bibinfo {author} {\bibfnamefont {W.~M.}\ \bibnamefont
  {Linhart}}, \bibinfo {author} {\bibfnamefont {M.}~\bibnamefont {Rybak}},
  \bibinfo {author} {\bibfnamefont {M.}~\bibnamefont {Birowska}}, \bibinfo
  {author} {\bibfnamefont {P.}~\bibnamefont {Scharoch}}, \bibinfo {author}
  {\bibfnamefont {K.}~\bibnamefont {Mosina}}, \bibinfo {author} {\bibfnamefont
  {V.}~\bibnamefont {Mazanek}}, \bibinfo {author} {\bibfnamefont
  {D.}~\bibnamefont {Kaczorowski}}, \bibinfo {author} {\bibfnamefont
  {Z.}~\bibnamefont {Sofer}},\ and\ \bibinfo {author} {\bibfnamefont
  {R.}~\bibnamefont {Kudrawiec}},\ }\href@noop {} {\bibfield  {journal}
  {\bibinfo  {journal} {J. Mater. Chem. C}\ }\textbf {\bibinfo {volume} {11}}
  (\bibinfo {year} {2023})}\BibitemShut {NoStop}%
\bibitem [{\citenamefont {Torres}\ \emph {et~al.}(2023)\citenamefont {Torres},
  \citenamefont {Kuc}, \citenamefont {Maschio}, \citenamefont {Pham},
  \citenamefont {Reidy}, \citenamefont {Dekanovsky}, \citenamefont {Sofer},
  \citenamefont {Ross}, ,\ and\ \citenamefont {Klein}}]{Torres2023}%
  \BibitemOpen
  \bibfield  {author} {\bibinfo {author} {\bibfnamefont {K.}~\bibnamefont
  {Torres}}, \bibinfo {author} {\bibfnamefont {A.}~\bibnamefont {Kuc}},
  \bibinfo {author} {\bibfnamefont {L.}~\bibnamefont {Maschio}}, \bibinfo
  {author} {\bibfnamefont {T.}~\bibnamefont {Pham}}, \bibinfo {author}
  {\bibfnamefont {K.}~\bibnamefont {Reidy}}, \bibinfo {author} {\bibfnamefont
  {L.}~\bibnamefont {Dekanovsky}}, \bibinfo {author} {\bibfnamefont
  {Z.}~\bibnamefont {Sofer}}, \bibinfo {author} {\bibfnamefont {F.~M.}\
  \bibnamefont {Ross}}, ,\ and\ \bibinfo {author} {\bibfnamefont
  {J.}~\bibnamefont {Klein}},\ }\href@noop {} {\bibfield  {journal} {\bibinfo
  {journal} {Adv. Funct. Mater.}\ }\textbf {\bibinfo {volume} {32}} (\bibinfo
  {year} {2023})}\BibitemShut {NoStop}%
\bibitem [{\citenamefont {Tschudin}\ \emph {et~al.}(2024)\citenamefont
  {Tschudin}, \citenamefont {Broadway}, \citenamefont {Siegwolf}, \citenamefont
  {Schrader}, \citenamefont {Telford}, \citenamefont {Gross}, \citenamefont
  {Cox}, \citenamefont {Dubois}, \citenamefont {Chica}, \citenamefont
  {Rama-Eiroa}, \citenamefont {J.~G.~Santos}, \citenamefont {Poggio},
  \citenamefont {Ziebel}, \citenamefont {Dean}, \citenamefont {Roy},\ and\
  \citenamefont {Maletinsky}}]{Tschudin2024}%
  \BibitemOpen
  \bibfield  {author} {\bibinfo {author} {\bibfnamefont {M.~A.}\ \bibnamefont
  {Tschudin}}, \bibinfo {author} {\bibfnamefont {D.~A.}\ \bibnamefont
  {Broadway}}, \bibinfo {author} {\bibfnamefont {P.}~\bibnamefont {Siegwolf}},
  \bibinfo {author} {\bibfnamefont {C.}~\bibnamefont {Schrader}}, \bibinfo
  {author} {\bibfnamefont {E.~J.}\ \bibnamefont {Telford}}, \bibinfo {author}
  {\bibfnamefont {B.}~\bibnamefont {Gross}}, \bibinfo {author} {\bibfnamefont
  {J.}~\bibnamefont {Cox}}, \bibinfo {author} {\bibfnamefont {A.~E.~E.}\
  \bibnamefont {Dubois}}, \bibinfo {author} {\bibfnamefont {D.~G.}\
  \bibnamefont {Chica}}, \bibinfo {author} {\bibfnamefont {R.}~\bibnamefont
  {Rama-Eiroa}}, \bibinfo {author} {\bibfnamefont {E.}~\bibnamefont
  {J.~G.~Santos}}, \bibinfo {author} {\bibfnamefont {M.}~\bibnamefont
  {Poggio}}, \bibinfo {author} {\bibfnamefont {M.~E.}\ \bibnamefont {Ziebel}},
  \bibinfo {author} {\bibfnamefont {C.~R.}\ \bibnamefont {Dean}}, \bibinfo
  {author} {\bibfnamefont {X.}~\bibnamefont {Roy}},\ and\ \bibinfo {author}
  {\bibfnamefont {P.}~\bibnamefont {Maletinsky}},\ }\href@noop {} {\bibfield
  {journal} {\bibinfo  {journal} {Nature Commun.}\ }\textbf {\bibinfo {volume}
  {15}},\ \bibinfo {pages} {6005} (\bibinfo {year} {2024})}\BibitemShut
  {NoStop}%
\bibitem [{\citenamefont {Göser}\ \emph {et~al.}(1990)\citenamefont {Göser},
  \citenamefont {Paul},\ and\ \citenamefont {Kahle}}]{Goser1990}%
  \BibitemOpen
  \bibfield  {author} {\bibinfo {author} {\bibfnamefont {O.}~\bibnamefont
  {Göser}}, \bibinfo {author} {\bibfnamefont {W.}~\bibnamefont {Paul}},\ and\
  \bibinfo {author} {\bibfnamefont {H.~G.}\ \bibnamefont {Kahle}},\ }\href@noop
  {} {\bibfield  {journal} {\bibinfo  {journal} {J. Magn. Magn. Mater.}\
  }\textbf {\bibinfo {volume} {92}} (\bibinfo {year} {1990})}\BibitemShut
  {NoStop}%
\bibitem [{\citenamefont {Lee}\ \emph {et~al.}(2016)\citenamefont {Lee},
  \citenamefont {Lee}, \citenamefont {Ryoo}, \citenamefont {Kang},
  \citenamefont {Kim}, \citenamefont {Kim}, \citenamefont {Park}, \citenamefont
  {Park},\ and\ \citenamefont {Cheong}}]{Lee2016}%
  \BibitemOpen
  \bibfield  {author} {\bibinfo {author} {\bibfnamefont {J.-U.}\ \bibnamefont
  {Lee}}, \bibinfo {author} {\bibfnamefont {S.}~\bibnamefont {Lee}}, \bibinfo
  {author} {\bibfnamefont {J.~H.}\ \bibnamefont {Ryoo}}, \bibinfo {author}
  {\bibfnamefont {S.}~\bibnamefont {Kang}}, \bibinfo {author} {\bibfnamefont
  {T.~Y.}\ \bibnamefont {Kim}}, \bibinfo {author} {\bibfnamefont
  {P.}~\bibnamefont {Kim}}, \bibinfo {author} {\bibfnamefont {C.-H.}\
  \bibnamefont {Park}}, \bibinfo {author} {\bibfnamefont {J.-G.}\ \bibnamefont
  {Park}},\ and\ \bibinfo {author} {\bibfnamefont {H.}~\bibnamefont {Cheong}},\
  }\href@noop {} {\bibfield  {journal} {\bibinfo  {journal} {Nano Lett.}\
  }\textbf {\bibinfo {volume} {16}} (\bibinfo {year} {2016})}\BibitemShut
  {NoStop}%
\bibitem [{\citenamefont {Kim}\ \emph {et~al.}(2016)\citenamefont {Kim},
  \citenamefont {Lim}, \citenamefont {Lee}, \citenamefont {Lee}, \citenamefont
  {Kim}, \citenamefont {Park}, \citenamefont {Jeon}, \citenamefont {Park},
  \citenamefont {Park},\ and\ \citenamefont {Cheong}}]{Kim2019a}%
  \BibitemOpen
  \bibfield  {author} {\bibinfo {author} {\bibfnamefont {K.}~\bibnamefont
  {Kim}}, \bibinfo {author} {\bibfnamefont {S.~Y.}\ \bibnamefont {Lim}},
  \bibinfo {author} {\bibfnamefont {J.-U.}\ \bibnamefont {Lee}}, \bibinfo
  {author} {\bibfnamefont {S.}~\bibnamefont {Lee}}, \bibinfo {author}
  {\bibfnamefont {T.~Y.}\ \bibnamefont {Kim}}, \bibinfo {author} {\bibfnamefont
  {K.}~\bibnamefont {Park}}, \bibinfo {author} {\bibfnamefont {G.~S.}\
  \bibnamefont {Jeon}}, \bibinfo {author} {\bibfnamefont {C.-H.}\ \bibnamefont
  {Park}}, \bibinfo {author} {\bibfnamefont {J.-G.}\ \bibnamefont {Park}},\
  and\ \bibinfo {author} {\bibfnamefont {H.}~\bibnamefont {Cheong}},\
  }\href@noop {} {\bibfield  {journal} {\bibinfo  {journal} {Nature Commun.}\
  }\textbf {\bibinfo {volume} {10}} (\bibinfo {year} {2016})}\BibitemShut
  {NoStop}%
\bibitem [{\citenamefont {Kim}\ \emph {et~al.}(2019)\citenamefont {Kim},
  \citenamefont {Lee},\ and\ \citenamefont {Cheong}}]{Kim2019b}%
  \BibitemOpen
  \bibfield  {author} {\bibinfo {author} {\bibfnamefont {K.}~\bibnamefont
  {Kim}}, \bibinfo {author} {\bibfnamefont {J.-U.}\ \bibnamefont {Lee}},\ and\
  \bibinfo {author} {\bibfnamefont {H.}~\bibnamefont {Cheong}},\ }\href@noop {}
  {\bibfield  {journal} {\bibinfo  {journal} {Nanotechnol.}\ }\textbf {\bibinfo
  {volume} {30}} (\bibinfo {year} {2019})}\BibitemShut {NoStop}%
\bibitem [{\citenamefont {Sun}\ \emph {et~al.}(2019)\citenamefont {Sun},
  \citenamefont {Pang},\ and\ \citenamefont {Zhang}}]{Sun2021}%
  \BibitemOpen
  \bibfield  {author} {\bibinfo {author} {\bibfnamefont {Y.-J.}\ \bibnamefont
  {Sun}}, \bibinfo {author} {\bibfnamefont {S.-M.}\ \bibnamefont {Pang}},\ and\
  \bibinfo {author} {\bibfnamefont {J.}~\bibnamefont {Zhang}},\ }\href@noop {}
  {\bibfield  {journal} {\bibinfo  {journal} {Chin. Phys. B}\ }\textbf
  {\bibinfo {volume} {30}} (\bibinfo {year} {2019})}\BibitemShut {NoStop}%
\bibitem [{\citenamefont {Zhang}\ \emph {et~al.}(2020)\citenamefont {Zhang},
  \citenamefont {Wu}, \citenamefont {Lyu}, \citenamefont {Wu}, \citenamefont
  {Zhao}, \citenamefont {Chen}, \citenamefont {Jia}, \citenamefont {Zhang},
  \citenamefont {Wang}, \citenamefont {Wang}, \citenamefont {Chen},
  \citenamefont {Mei}, \citenamefont {Taniguchi}, \citenamefont {Watanabe},
  \citenamefont {Yan}, \citenamefont {Liu}, \citenamefont {Huang},
  \citenamefont {Zhao},\ and\ \citenamefont {Huang}}]{Zhang2020}%
  \BibitemOpen
  \bibfield  {author} {\bibinfo {author} {\bibfnamefont {Y.}~\bibnamefont
  {Zhang}}, \bibinfo {author} {\bibfnamefont {X.}~\bibnamefont {Wu}}, \bibinfo
  {author} {\bibfnamefont {B.}~\bibnamefont {Lyu}}, \bibinfo {author}
  {\bibfnamefont {M.}~\bibnamefont {Wu}}, \bibinfo {author} {\bibfnamefont
  {S.}~\bibnamefont {Zhao}}, \bibinfo {author} {\bibfnamefont {J.}~\bibnamefont
  {Chen}}, \bibinfo {author} {\bibfnamefont {M.}~\bibnamefont {Jia}}, \bibinfo
  {author} {\bibfnamefont {C.}~\bibnamefont {Zhang}}, \bibinfo {author}
  {\bibfnamefont {L.}~\bibnamefont {Wang}}, \bibinfo {author} {\bibfnamefont
  {X.}~\bibnamefont {Wang}}, \bibinfo {author} {\bibfnamefont {Y.}~\bibnamefont
  {Chen}}, \bibinfo {author} {\bibfnamefont {J.}~\bibnamefont {Mei}}, \bibinfo
  {author} {\bibfnamefont {T.}~\bibnamefont {Taniguchi}}, \bibinfo {author}
  {\bibfnamefont {K.}~\bibnamefont {Watanabe}}, \bibinfo {author}
  {\bibfnamefont {H.}~\bibnamefont {Yan}}, \bibinfo {author} {\bibfnamefont
  {Q.}~\bibnamefont {Liu}}, \bibinfo {author} {\bibfnamefont {L.}~\bibnamefont
  {Huang}}, \bibinfo {author} {\bibfnamefont {Y.}~\bibnamefont {Zhao}},\ and\
  \bibinfo {author} {\bibfnamefont {M.}~\bibnamefont {Huang}},\ }\href@noop {}
  {\bibfield  {journal} {\bibinfo  {journal} {Nano Lett.}\ }\textbf {\bibinfo
  {volume} {20}},\ \bibinfo {pages} {729} (\bibinfo {year} {2020})}\BibitemShut
  {NoStop}%
\bibitem [{\citenamefont {McCreary}\ \emph
  {et~al.}(2020{\natexlab{a}})\citenamefont {McCreary}, \citenamefont
  {Simpson}, \citenamefont {Mai}, \citenamefont {McMichael}, \citenamefont
  {Douglas}, \citenamefont {Butch}, \citenamefont {Dennis}, \citenamefont
  {Vald\'es~Aguilar},\ and\ \citenamefont {Hight~Walker}}]{McCreary2020a}%
  \BibitemOpen
  \bibfield  {author} {\bibinfo {author} {\bibfnamefont {A.}~\bibnamefont
  {McCreary}}, \bibinfo {author} {\bibfnamefont {J.~R.}\ \bibnamefont
  {Simpson}}, \bibinfo {author} {\bibfnamefont {T.~T.}\ \bibnamefont {Mai}},
  \bibinfo {author} {\bibfnamefont {R.~D.}\ \bibnamefont {McMichael}}, \bibinfo
  {author} {\bibfnamefont {J.~E.}\ \bibnamefont {Douglas}}, \bibinfo {author}
  {\bibfnamefont {N.}~\bibnamefont {Butch}}, \bibinfo {author} {\bibfnamefont
  {C.}~\bibnamefont {Dennis}}, \bibinfo {author} {\bibfnamefont
  {R.}~\bibnamefont {Vald\'es~Aguilar}},\ and\ \bibinfo {author} {\bibfnamefont
  {A.~R.}\ \bibnamefont {Hight~Walker}},\ }\href@noop {} {\bibfield  {journal}
  {\bibinfo  {journal} {Phys. Rev. B}\ }\textbf {\bibinfo {volume} {101}},\
  \bibinfo {pages} {064416} (\bibinfo {year} {2020}{\natexlab{a}})}\BibitemShut
  {NoStop}%
\bibitem [{\citenamefont {McCreary}\ \emph
  {et~al.}(2020{\natexlab{b}})\citenamefont {McCreary}, \citenamefont {Mai},
  \citenamefont {Utermohlen}, \citenamefont {Simpson}, \citenamefont {Garrity},
  \citenamefont {Feng}, \citenamefont {Shcherbakov}, \citenamefont {Zhu},
  \citenamefont {Hu}, \citenamefont {Weber}, \citenamefont {Watanabe},
  \citenamefont {Taniguchi}, \citenamefont {Goldberger}, \citenamefont {Mao},
  \citenamefont {Lau}, \citenamefont {Lu2}, \citenamefont {Trivedi},
  \citenamefont {Aguilar},\ and\ \citenamefont {Walker}}]{McCreary2020b}%
  \BibitemOpen
  \bibfield  {author} {\bibinfo {author} {\bibfnamefont {A.}~\bibnamefont
  {McCreary}}, \bibinfo {author} {\bibfnamefont {T.~T.}\ \bibnamefont {Mai}},
  \bibinfo {author} {\bibfnamefont {F.~G.}\ \bibnamefont {Utermohlen}},
  \bibinfo {author} {\bibfnamefont {J.~R.}\ \bibnamefont {Simpson}}, \bibinfo
  {author} {\bibfnamefont {K.~F.}\ \bibnamefont {Garrity}}, \bibinfo {author}
  {\bibfnamefont {X.}~\bibnamefont {Feng}}, \bibinfo {author} {\bibfnamefont
  {D.}~\bibnamefont {Shcherbakov}}, \bibinfo {author} {\bibfnamefont
  {Y.}~\bibnamefont {Zhu}}, \bibinfo {author} {\bibfnamefont {J.}~\bibnamefont
  {Hu}}, \bibinfo {author} {\bibfnamefont {D.}~\bibnamefont {Weber}}, \bibinfo
  {author} {\bibfnamefont {K.}~\bibnamefont {Watanabe}}, \bibinfo {author}
  {\bibfnamefont {T.}~\bibnamefont {Taniguchi}}, \bibinfo {author}
  {\bibfnamefont {J.~E.}\ \bibnamefont {Goldberger}}, \bibinfo {author}
  {\bibfnamefont {Z.}~\bibnamefont {Mao}}, \bibinfo {author} {\bibfnamefont
  {C.~N.}\ \bibnamefont {Lau}}, \bibinfo {author} {\bibfnamefont
  {Y.}~\bibnamefont {Lu2}}, \bibinfo {author} {\bibfnamefont {N.}~\bibnamefont
  {Trivedi}}, \bibinfo {author} {\bibfnamefont {R.~V.}\ \bibnamefont
  {Aguilar}},\ and\ \bibinfo {author} {\bibfnamefont {A.~R.~H.}\ \bibnamefont
  {Walker}},\ }\href@noop {} {\bibfield  {journal} {\bibinfo  {journal} {Nature
  Commun.}\ }\textbf {\bibinfo {volume} {11}},\ \bibinfo {pages} {3879}
  (\bibinfo {year} {2020}{\natexlab{b}})}\BibitemShut {NoStop}%
\bibitem [{\citenamefont {Kong}\ and\ \citenamefont {Liang}(2024)}]{Kong2024}%
  \BibitemOpen
  \bibfield  {author} {\bibinfo {author} {\bibfnamefont {X.}~\bibnamefont
  {Kong}}\ and\ \bibinfo {author} {\bibfnamefont {P.~G.~L.}\ \bibnamefont
  {Liang}},\ }\href@noop {} {\bibfield  {journal} {\bibinfo  {journal} {npj 2D
  mater. appl.}\ }\textbf {\bibinfo {volume} {8}},\ \bibinfo {pages} {82}
  (\bibinfo {year} {2024})}\BibitemShut {NoStop}%
\bibitem [{\citenamefont {Staros}\ \emph {et~al.}(2022)\citenamefont {Staros},
  \citenamefont {Hu}, \citenamefont {Tiihonen}, \citenamefont {Nanguneri},
  \citenamefont {Krogel}, \citenamefont {Bennett}, \citenamefont {Heinonen},
  \citenamefont {Ganesh},\ and\ \citenamefont {Rubenstein}}]{Staros2022}%
  \BibitemOpen
  \bibfield  {author} {\bibinfo {author} {\bibfnamefont {D.}~\bibnamefont
  {Staros}}, \bibinfo {author} {\bibfnamefont {G.}~\bibnamefont {Hu}}, \bibinfo
  {author} {\bibfnamefont {J.}~\bibnamefont {Tiihonen}}, \bibinfo {author}
  {\bibfnamefont {R.}~\bibnamefont {Nanguneri}}, \bibinfo {author}
  {\bibfnamefont {J.}~\bibnamefont {Krogel}}, \bibinfo {author} {\bibfnamefont
  {M.~C.}\ \bibnamefont {Bennett}}, \bibinfo {author} {\bibfnamefont
  {O.}~\bibnamefont {Heinonen}}, \bibinfo {author} {\bibfnamefont
  {P.}~\bibnamefont {Ganesh}},\ and\ \bibinfo {author} {\bibfnamefont
  {B.}~\bibnamefont {Rubenstein}},\ }\href@noop {} {\bibfield  {journal}
  {\bibinfo  {journal} {J. Chem Phys.}\ }\textbf {\bibinfo {volume} {156}},\
  \bibinfo {pages} {014707} (\bibinfo {year} {2022})}\BibitemShut {NoStop}%
\bibitem [{\citenamefont {Kresse}\ and\ \citenamefont {Hafner}(1993)}]{vasp1}%
  \BibitemOpen
  \bibfield  {author} {\bibinfo {author} {\bibfnamefont {G.}~\bibnamefont
  {Kresse}}\ and\ \bibinfo {author} {\bibfnamefont {J.}~\bibnamefont
  {Hafner}},\ }\href@noop {} {\bibfield  {journal} {\bibinfo  {journal} {Phys.
  Rev. B}\ }\textbf {\bibinfo {volume} {47}},\ \bibinfo {pages} {558} (\bibinfo
  {year} {1993})}\BibitemShut {NoStop}%
\bibitem [{\citenamefont {Kresse}\ and\ \citenamefont
  {Furthmüller}(1996)}]{vasp2}%
  \BibitemOpen
  \bibfield  {author} {\bibinfo {author} {\bibfnamefont {G.}~\bibnamefont
  {Kresse}}\ and\ \bibinfo {author} {\bibfnamefont {J.}~\bibnamefont
  {Furthmüller}},\ }\href@noop {} {\bibfield  {journal} {\bibinfo  {journal}
  {Comput. Mater. Sci.}\ }\textbf {\bibinfo {volume} {6}},\ \bibinfo {pages}
  {15} (\bibinfo {year} {1996})}\BibitemShut {NoStop}%
\bibitem [{\citenamefont {Perdew}\ \emph {et~al.}(1996)\citenamefont {Perdew},
  \citenamefont {Burke},\ and\ \citenamefont {Ernzerhof}}]{pbe1}%
  \BibitemOpen
  \bibfield  {author} {\bibinfo {author} {\bibfnamefont {J.~P.}\ \bibnamefont
  {Perdew}}, \bibinfo {author} {\bibfnamefont {K.}~\bibnamefont {Burke}},\ and\
  \bibinfo {author} {\bibfnamefont {M.}~\bibnamefont {Ernzerhof}},\ }\href@noop
  {} {\bibfield  {journal} {\bibinfo  {journal} {Phys. Rev. Lett.}\ }\textbf
  {\bibinfo {volume} {77}},\ \bibinfo {pages} {3865} (\bibinfo {year}
  {1996})}\BibitemShut {NoStop}%
\bibitem [{\citenamefont {Blöchl}(1994)}]{paw1}%
  \BibitemOpen
  \bibfield  {author} {\bibinfo {author} {\bibfnamefont {P.~E.}\ \bibnamefont
  {Blöchl}},\ }\href@noop {} {\bibfield  {journal} {\bibinfo  {journal} {Phys.
  Rev. B}\ }\textbf {\bibinfo {volume} {5}},\ \bibinfo {pages} {17953}
  (\bibinfo {year} {1994})}\BibitemShut {NoStop}%
\bibitem [{\citenamefont {Kresse}\ and\ \citenamefont {Joubert}(1999)}]{paw2}%
  \BibitemOpen
  \bibfield  {author} {\bibinfo {author} {\bibfnamefont {G.}~\bibnamefont
  {Kresse}}\ and\ \bibinfo {author} {\bibfnamefont {D.}~\bibnamefont
  {Joubert}},\ }\href@noop {} {\bibfield  {journal} {\bibinfo  {journal} {Phys.
  Rev. B}\ }\textbf {\bibinfo {volume} {59}},\ \bibinfo {pages} {1758}
  (\bibinfo {year} {1999})}\BibitemShut {NoStop}%
\bibitem [{\citenamefont {Dudarev}\ \emph {et~al.}(1998)\citenamefont
  {Dudarev}, \citenamefont {Botton}, \citenamefont {Savrasov}, \citenamefont
  {Humphreys},\ and\ \citenamefont {Sutton}}]{dudarev98}%
  \BibitemOpen
  \bibfield  {author} {\bibinfo {author} {\bibfnamefont {S.~L.}\ \bibnamefont
  {Dudarev}}, \bibinfo {author} {\bibfnamefont {G.~A.}\ \bibnamefont {Botton}},
  \bibinfo {author} {\bibfnamefont {S.~Y.}\ \bibnamefont {Savrasov}}, \bibinfo
  {author} {\bibfnamefont {C.~J.}\ \bibnamefont {Humphreys}},\ and\ \bibinfo
  {author} {\bibfnamefont {A.~P.}\ \bibnamefont {Sutton}},\ }\href@noop {}
  {\bibfield  {journal} {\bibinfo  {journal} {Phys. Rev. B}\ }\textbf {\bibinfo
  {volume} {57}},\ \bibinfo {pages} {1505} (\bibinfo {year}
  {1998})}\BibitemShut {NoStop}%
\bibitem [{\citenamefont {Grimme}\ \emph {et~al.}(2010)\citenamefont {Grimme},
  \citenamefont {Antony}, \citenamefont {Ehrlich},\ and\ \citenamefont
  {Krieg}}]{vdw1}%
  \BibitemOpen
  \bibfield  {author} {\bibinfo {author} {\bibfnamefont {S.}~\bibnamefont
  {Grimme}}, \bibinfo {author} {\bibfnamefont {J.}~\bibnamefont {Antony}},
  \bibinfo {author} {\bibfnamefont {S.}~\bibnamefont {Ehrlich}},\ and\ \bibinfo
  {author} {\bibfnamefont {S.}~\bibnamefont {Krieg}},\ }\href@noop {}
  {\bibfield  {journal} {\bibinfo  {journal} {J. Chem. Phys.}\ }\textbf
  {\bibinfo {volume} {132}},\ \bibinfo {pages} {154104} (\bibinfo {year}
  {2010})}\BibitemShut {NoStop}%
\bibitem [{\citenamefont {Grimme}\ \emph {et~al.}(2011)\citenamefont {Grimme},
  \citenamefont {Ehrlich},\ and\ \citenamefont {Goerigk}}]{vdw2}%
  \BibitemOpen
  \bibfield  {author} {\bibinfo {author} {\bibfnamefont {S.}~\bibnamefont
  {Grimme}}, \bibinfo {author} {\bibfnamefont {S.}~\bibnamefont {Ehrlich}},\
  and\ \bibinfo {author} {\bibfnamefont {L.}~\bibnamefont {Goerigk}},\
  }\href@noop {} {\bibfield  {journal} {\bibinfo  {journal} {J. Comput. Chem.}\
  }\textbf {\bibinfo {volume} {32}},\ \bibinfo {pages} {1456} (\bibinfo {year}
  {2011})}\BibitemShut {NoStop}%
\bibitem [{\citenamefont {Parlinski}\ \emph {et~al.}(1997)\citenamefont
  {Parlinski}, \citenamefont {Li},\ and\ \citenamefont
  {Kawazoe}}]{parlinski97}%
  \BibitemOpen
  \bibfield  {author} {\bibinfo {author} {\bibfnamefont {K.}~\bibnamefont
  {Parlinski}}, \bibinfo {author} {\bibfnamefont {Z.-Q.}\ \bibnamefont {Li}},\
  and\ \bibinfo {author} {\bibfnamefont {Y.}~\bibnamefont {Kawazoe}},\
  }\href@noop {} {\bibfield  {journal} {\bibinfo  {journal} {Phys. Rev. Lett.}\
  }\textbf {\bibinfo {volume} {78}},\ \bibinfo {pages} {4063} (\bibinfo {year}
  {1997})}\BibitemShut {NoStop}%
\bibitem [{\citenamefont {Togo}\ and\ \citenamefont {Tanaka}(2015)}]{togo15}%
  \BibitemOpen
  \bibfield  {author} {\bibinfo {author} {\bibfnamefont {A.}~\bibnamefont
  {Togo}}\ and\ \bibinfo {author} {\bibfnamefont {I.}~\bibnamefont {Tanaka}},\
  }\href@noop {} {\bibfield  {journal} {\bibinfo  {journal} {Scripta. Mater.}\
  }\textbf {\bibinfo {volume} {108}},\ \bibinfo {pages} {1} (\bibinfo {year}
  {2015})}\BibitemShut {NoStop}%
\bibitem [{\citenamefont {Cardona}(1983)}]{cardona83}%
  \BibitemOpen
  \bibinfo {editor} {\bibfnamefont {M.}~\bibnamefont {Cardona}},\ ed.,\
  \href@noop {} {\emph {\bibinfo {title} {Light Scattering in Solids I}}},\
  Vol.~\bibinfo {volume} {8}\ (\bibinfo  {publisher} {Springer--Verlag},\
  \bibinfo {address} {Berlin Heidelberg},\ \bibinfo {year} {1983})\BibitemShut
  {NoStop}%
\bibitem [{\citenamefont {Umari}\ \emph {et~al.}(2001)\citenamefont {Umari},
  \citenamefont {Pasquarello},\ and\ \citenamefont {Corso}}]{umari01}%
  \BibitemOpen
  \bibfield  {author} {\bibinfo {author} {\bibfnamefont {P.}~\bibnamefont
  {Umari}}, \bibinfo {author} {\bibfnamefont {A.}~\bibnamefont {Pasquarello}},\
  and\ \bibinfo {author} {\bibfnamefont {A.~D.}\ \bibnamefont {Corso}},\
  }\href@noop {} {\bibfield  {journal} {\bibinfo  {journal} {Phys. Rev. B}\
  }\textbf {\bibinfo {volume} {63}},\ \bibinfo {pages} {094305} (\bibinfo
  {year} {2001})}\BibitemShut {NoStop}%
\bibitem [{\citenamefont {Umari}\ \emph {et~al.}(2003)\citenamefont {Umari},
  \citenamefont {Gonze},\ and\ \citenamefont {Pasquarello}}]{umari03}%
  \BibitemOpen
  \bibfield  {author} {\bibinfo {author} {\bibfnamefont {P.}~\bibnamefont
  {Umari}}, \bibinfo {author} {\bibfnamefont {X.}~\bibnamefont {Gonze}},\ and\
  \bibinfo {author} {\bibfnamefont {A.}~\bibnamefont {Pasquarello}},\
  }\href@noop {} {\bibfield  {journal} {\bibinfo  {journal} {Phys. Rev. Lett.}\
  }\textbf {\bibinfo {volume} {90}},\ \bibinfo {pages} {027401} (\bibinfo
  {year} {2003})}\BibitemShut {NoStop}%
\bibitem [{\citenamefont {Sternik}\ and\ \citenamefont
  {Wdowik}(2018)}]{sternik18}%
  \BibitemOpen
  \bibfield  {author} {\bibinfo {author} {\bibfnamefont {M.}~\bibnamefont
  {Sternik}}\ and\ \bibinfo {author} {\bibfnamefont {U.~D.}\ \bibnamefont
  {Wdowik}},\ }\href@noop {} {\bibfield  {journal} {\bibinfo  {journal} {Phys.
  Chem. Chem. Phys}\ }\textbf {\bibinfo {volume} {20}},\ \bibinfo {pages}
  {7754} (\bibinfo {year} {2018})}\BibitemShut {NoStop}%
\bibitem [{\citenamefont {Calizo}\ \emph {et~al.}(2007)\citenamefont {Calizo},
  \citenamefont {Balandin}, \citenamefont {Bao}, \citenamefont {Miao},\ and\
  \citenamefont {Lau}}]{calizo07}%
  \BibitemOpen
  \bibfield  {author} {\bibinfo {author} {\bibfnamefont {I.}~\bibnamefont
  {Calizo}}, \bibinfo {author} {\bibfnamefont {A.~A.}\ \bibnamefont
  {Balandin}}, \bibinfo {author} {\bibfnamefont {W.}~\bibnamefont {Bao}},
  \bibinfo {author} {\bibfnamefont {F.}~\bibnamefont {Miao}},\ and\ \bibinfo
  {author} {\bibfnamefont {C.~N.}\ \bibnamefont {Lau}},\ }\href@noop {}
  {\bibfield  {journal} {\bibinfo  {journal} {Nano Lett.}\ }\textbf {\bibinfo
  {volume} {7}},\ \bibinfo {pages} {2645} (\bibinfo {year} {2007})}\BibitemShut
  {NoStop}%
\bibitem [{\citenamefont {Srivastava}\ and\ \citenamefont
  {Thomas}(2018)}]{srivastava18}%
  \BibitemOpen
  \bibfield  {author} {\bibinfo {author} {\bibfnamefont {G.~P.}\ \bibnamefont
  {Srivastava}}\ and\ \bibinfo {author} {\bibfnamefont {I.~O.}\ \bibnamefont
  {Thomas}},\ }\href@noop {} {\bibfield  {journal} {\bibinfo  {journal} {Phys.
  Rev. B}\ }\textbf {\bibinfo {volume} {98}},\ \bibinfo {pages} {035430}
  (\bibinfo {year} {2018})}\BibitemShut {NoStop}%
\bibitem [{\citenamefont {Tian}\ \emph {et~al.}(2016)\citenamefont {Tian},
  \citenamefont {Gray}, \citenamefont {Ji}, \citenamefont {Cava},\ and\
  \citenamefont {Burch}}]{tian16}%
  \BibitemOpen
  \bibfield  {author} {\bibinfo {author} {\bibfnamefont {Y.}~\bibnamefont
  {Tian}}, \bibinfo {author} {\bibfnamefont {M.~J.}\ \bibnamefont {Gray}},
  \bibinfo {author} {\bibfnamefont {H.}~\bibnamefont {Ji}}, \bibinfo {author}
  {\bibfnamefont {R.~J.}\ \bibnamefont {Cava}},\ and\ \bibinfo {author}
  {\bibfnamefont {K.~S.}\ \bibnamefont {Burch}},\ }\href@noop {} {\bibfield
  {journal} {\bibinfo  {journal} {2D Mater.}\ }\textbf {\bibinfo {volume}
  {3}},\ \bibinfo {pages} {025035} (\bibinfo {year} {2016})}\BibitemShut
  {NoStop}%
\bibitem [{\citenamefont {Ghosh}\ \emph {et~al.}(2021)\citenamefont {Ghosh},
  \citenamefont {Palit}, \citenamefont {Maity}, \citenamefont {Dwij},
  \citenamefont {Rana},\ and\ \citenamefont {Datta}}]{ghosh21}%
  \BibitemOpen
  \bibfield  {author} {\bibinfo {author} {\bibfnamefont {A.}~\bibnamefont
  {Ghosh}}, \bibinfo {author} {\bibfnamefont {M.}~\bibnamefont {Palit}},
  \bibinfo {author} {\bibfnamefont {S.}~\bibnamefont {Maity}}, \bibinfo
  {author} {\bibfnamefont {V.}~\bibnamefont {Dwij}}, \bibinfo {author}
  {\bibfnamefont {S.}~\bibnamefont {Rana}},\ and\ \bibinfo {author}
  {\bibfnamefont {S.}~\bibnamefont {Datta}},\ }\href@noop {} {\bibfield
  {journal} {\bibinfo  {journal} {Phys. Rev. B}\ }\textbf {\bibinfo {volume}
  {103}},\ \bibinfo {pages} {064431} (\bibinfo {year} {2021})}\BibitemShut
  {NoStop}%
\bibitem [{\citenamefont {Vaclavkova}\ \emph {et~al.}(2020)\citenamefont
  {Vaclavkova}, \citenamefont {Delhomme}, \citenamefont {Faugeras},
  \citenamefont {Potemski}, \citenamefont {Bogucki}, \citenamefont
  {Suffczyński}, \citenamefont {Kossacki}, \citenamefont {Wildes},
  \citenamefont {Grémaud},\ and\ \citenamefont {Saúl}}]{vaclavkova20}%
  \BibitemOpen
  \bibfield  {author} {\bibinfo {author} {\bibfnamefont {D.}~\bibnamefont
  {Vaclavkova}}, \bibinfo {author} {\bibfnamefont {A.}~\bibnamefont
  {Delhomme}}, \bibinfo {author} {\bibfnamefont {C.}~\bibnamefont {Faugeras}},
  \bibinfo {author} {\bibfnamefont {M.}~\bibnamefont {Potemski}}, \bibinfo
  {author} {\bibfnamefont {A.}~\bibnamefont {Bogucki}}, \bibinfo {author}
  {\bibfnamefont {J.}~\bibnamefont {Suffczyński}}, \bibinfo {author}
  {\bibfnamefont {P.}~\bibnamefont {Kossacki}}, \bibinfo {author}
  {\bibfnamefont {A.~R.}\ \bibnamefont {Wildes}}, \bibinfo {author}
  {\bibfnamefont {B.}~\bibnamefont {Grémaud}},\ and\ \bibinfo {author}
  {\bibfnamefont {A.}~\bibnamefont {Saúl}},\ }\href@noop {} {\bibfield
  {journal} {\bibinfo  {journal} {2D Mater.}\ }\textbf {\bibinfo {volume}
  {7}},\ \bibinfo {pages} {035030} (\bibinfo {year} {2020})}\BibitemShut
  {NoStop}%
\end{thebibliography}%
\newpage
\onecolumngrid
\clearpage
\appendix
\setcounter{figure}{0}
\setcounter{table}{0}
\section{Supplementary Information}
\label{appendix}
\setcounter{section}{0}
\makeatletter
\renewcommand{\theequation}{A\arabic{equation}}
\renewcommand{\thefigure}{A\arabic{figure}}
\renewcommand{\thetable}{A\arabic{table}}

\begin{table*}[h]
 \caption{Calculated and experimental \cite{lopez22,Lee2021} structural parameters of bulk CrSBr. Note: there are negligible difference in calculated structural parameters between AF-a, AF-b, and AF-c structures.}
\label{tab:structure}
    \centering
    \begin{tabular}{lcccccc}
    \hline \hline
     &a (\AA) &b (\AA) &c (\AA) &z$_{\mathrm{Cr}}$ &z$_{\mathrm{S}}$ &z$_{\mathrm{Br}}$ \\
     \hline
     Calc.                        &3.5381 &4.7459 &7.8514 &0.1297 &0.0765 &0.3562 \\
     Exp. \cite{lopez22} (1.8 K)  &3.1270 &4.7458 &7.9131 &0.1269 &0.0739 &0.3535 \\
     Exp. \cite{Lee2021} (100 K)    &3.5043 &4.7379 &7.9069 &       &       &       \\
    \hline \hline                
    \end{tabular}
\end{table*}

\begin{figure*}
    \centering
    \includegraphics[width=0.9\textwidth]{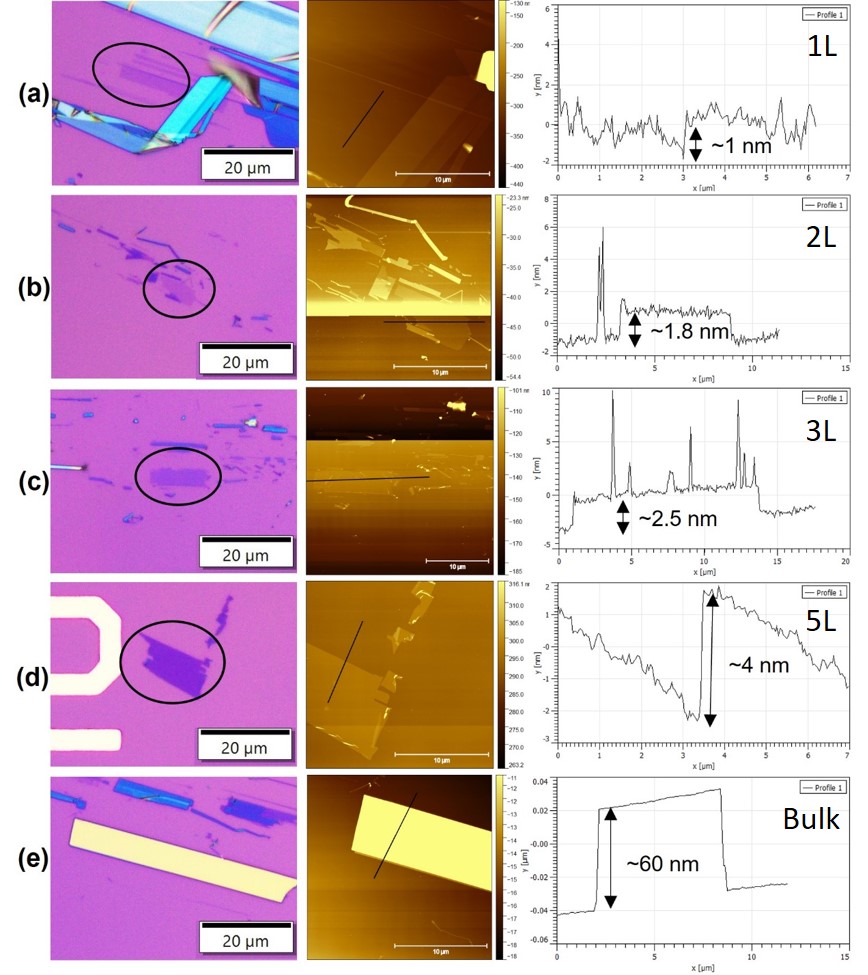}
    \caption{The optical microscope images, AFM images, and thickness profiles of 1L to 5L and bulk CrSBr. The profiles are determined along the line marked in black. The number of layers has been estimated using the thickness of 1L of about 0.8~nm \cite{Lee2021}. The lower-frequency phonon band ($\omega<150$ cm$^{-1}$ is dominated by Br vibrations. ).}
    \label{fig:AFM&profiles}
\end{figure*}

\begin{figure*}
    \centering
    \includegraphics[width=0.4\textwidth]{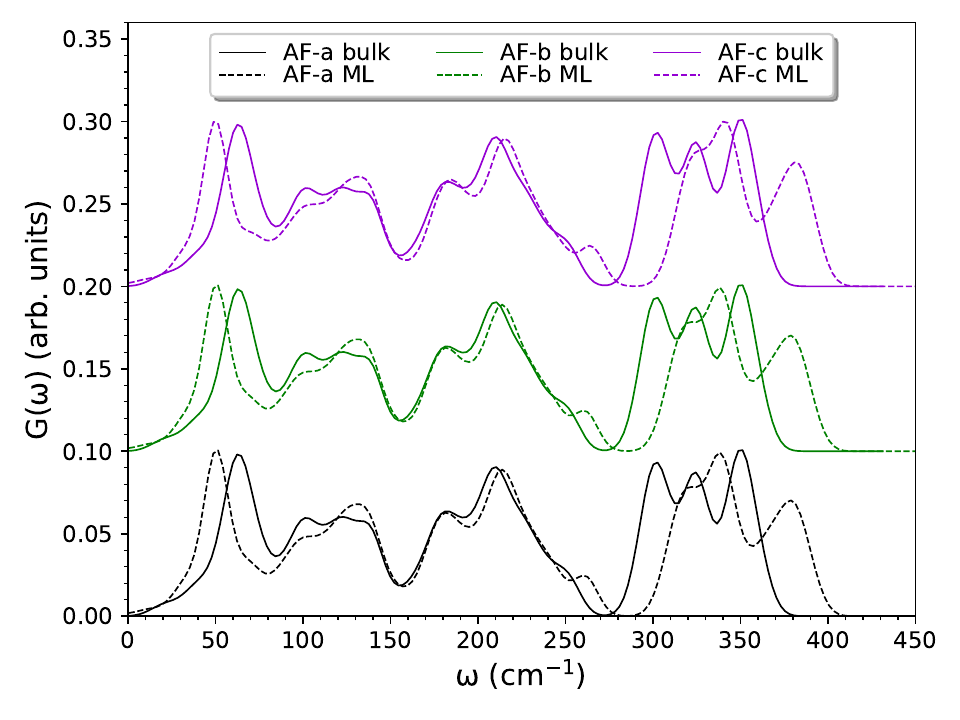}
    \includegraphics[width=0.4\textwidth]{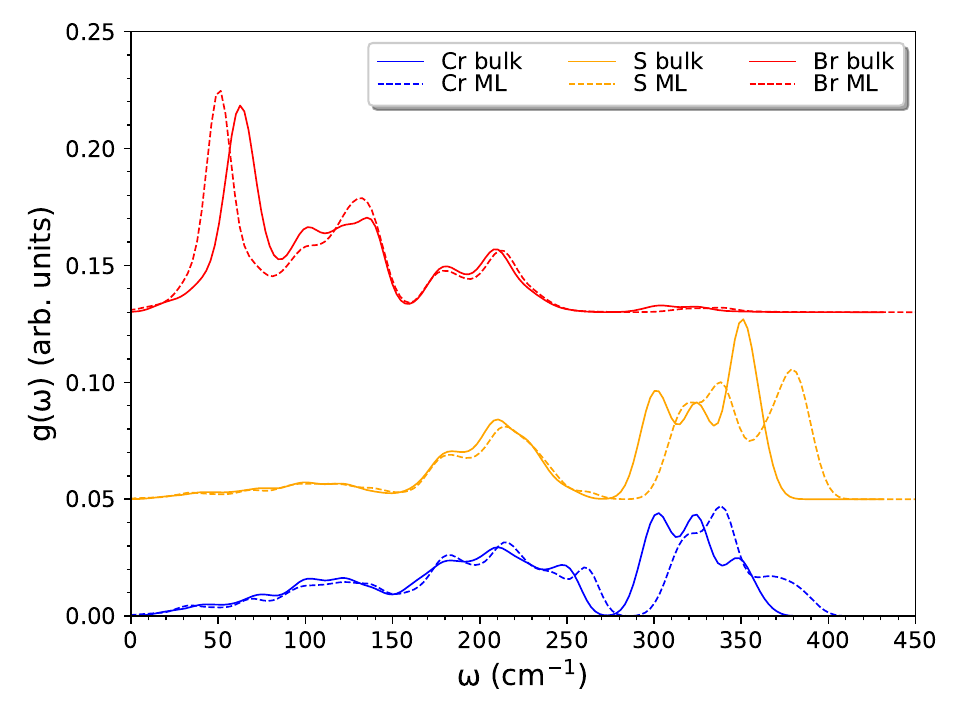}
    \caption{Total (left) and partial (right) phonon densities of states calculated for the AF-a, AF-b, and AF-c configurations of bulk (solid lines) and ML (dashed lines) CrSBr. Due to similarities of of the phonon spectra, the atomic resolved partial densities of phonon states are given for the AF-b configuration. The low-frequency phonon band ($\omega<150$ cm$^{-1}$) is dominated by vibrations of heavy Br atoms. Shared vibrations of Cr, S, and Br sublattices are observed in the medium-frequency phonon band ($150<\omega<280$ cm$^{-1}$), whereas shared vibrations of Cr and S atoms contribute most significantly to the high-frequency phonon band ($\omega>280$ cm$^{-1}$).}
    \label{fig:phDOS}
\end{figure*}

\begin{table*}[h]
 \caption{Atoms and their Wyckoff positions, contribution to the $\Gamma$-point phonons from each site, Raman tensors, and selection rules for CrSBr with \textit{Pmmn} space group (No. 59).}
    \label{tab:irreps}
    \centering
    \begin{tabular}{lcc}
    \hline \hline \\
    Atom &Wyckoff position &Irreducible representations \\
    \hline \\
    Cr &\textit{2b} &$A_g+B_{1u}+B_{2g}+B_{2u}+B_{3g}+B_{3u}$ \\
    S  &\textit{2a} &$A_g+B_{1u}+B_{2g}+B_{2u}+B_{3g}+B_{3u}$ \\
    Br &\textit{2a} &$A_g+B_{1u}+B_{2g}+B_{2u}+B_{3g}+B_{3u}$\\ 
    \hline
    \multicolumn{3}{c}{Raman tensors} \\
    $\mathbf{R}_{A_g} = \begin{pmatrix}
    a &0 &0 \\
    0 &b &0 \\
    0 &0 &c
    \end{pmatrix}$
    & $\mathbf{R}_{B_{2g}} = \begin{pmatrix}
    0 &0 &e \\
    0 &0 &0 \\
    e &0 &0
    \end{pmatrix}$
    &  $\mathbf{R}_{B_{3g}} = \begin{pmatrix}
    0 &0 &0 \\
    0 &0 &f \\
    0 &f &0
    \end{pmatrix}$ \\
    \hline
    \multicolumn{3}{c}{Activity and selection rules} \\
    \multicolumn{3}{c}{$\Gamma_{\mathrm{Raman}}=3A_g(\alpha_{xx},\alpha_{yy},\alpha_{zz})+3B_{2g}(\alpha_{xz})+3B_{3g}(\alpha_{yz})$} \\
    \multicolumn{3}{c}{$\Gamma_{\mathrm{IR}}=2B_{1u}(\mathbf{E} \parallel \mathbf{z})+2B_{2u}(\mathbf{E} \parallel \mathbf{y})+2B_{3u}(\mathbf{E} \parallel \mathbf{x})$} \\
    \multicolumn{3}{c}{$\Gamma_{\mathrm{acoustic}}=B_{1u}+B_{2u}+B_{3u}$} \\
    \hline \hline                
    \end{tabular}
\end{table*}

\begin{figure*}
    \centering
    \includegraphics[width=0.95\linewidth]{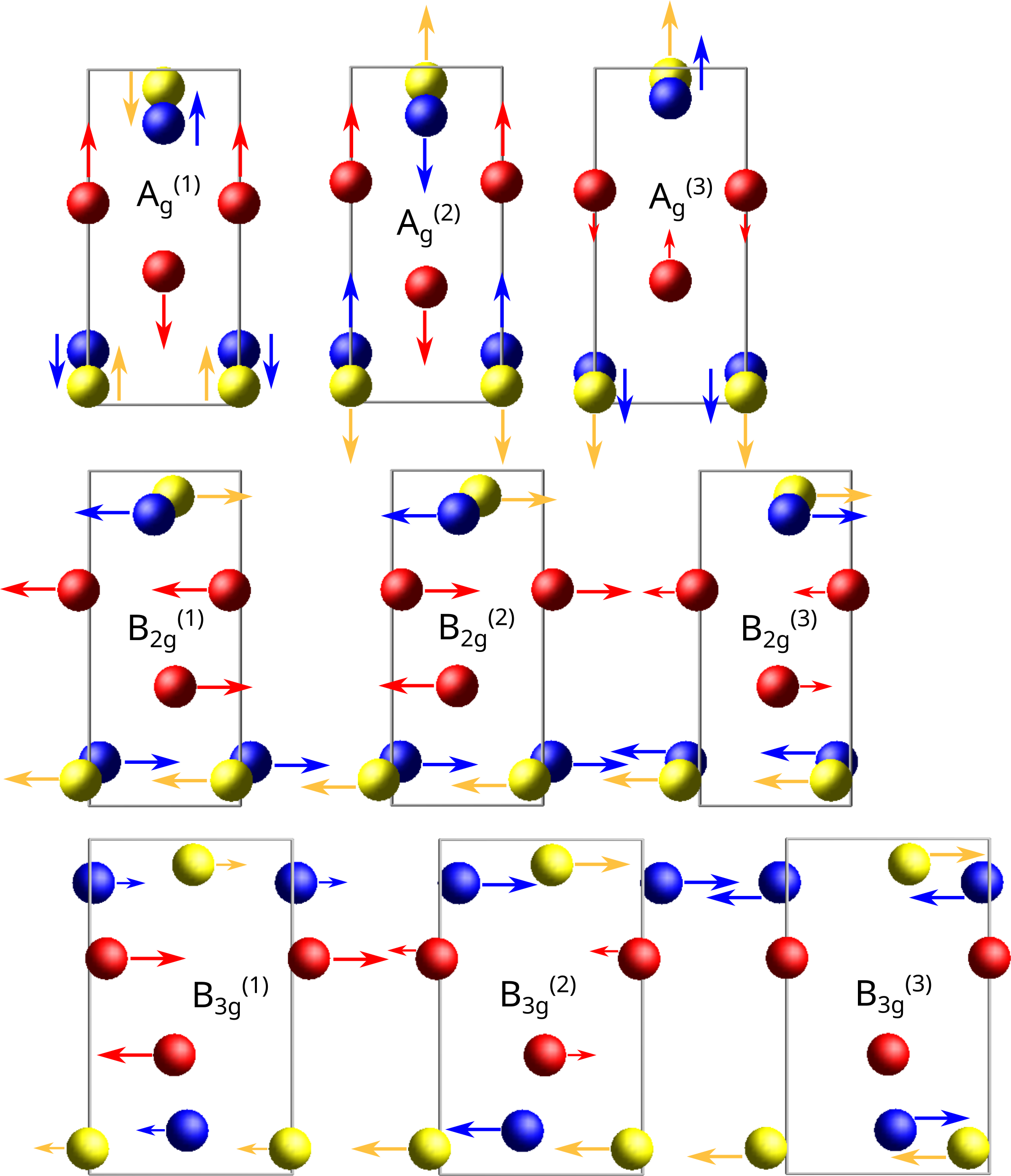}  
    \caption{Graphical representation of the Raman-active phonon modes in CrSBr. Atomic displacements are projected onto $(xz)$ plane for $A_g$ and $B_{2g}$ modes, whereas projection onto $(yz)$ plane is applied to show $B_{3g}$ modes. Modes are ordered with respect to their increasing frequencies.}
    \label{fig:atomic_displacements}
\end{figure*}

\begin{table*}[h]
 \caption{Calculated frequencies of the optical $\Gamma$-point phonon modes in bulk and monolayer (ML) CrSBr with different spatial spin arrangements (AF-a, AF-b, and AF-c). Frequencies of the Raman-active phonon modes are compared to those determined in the present and other experiments \cite{pawbake23,Klein23b}. Frequency unit: $cm^{-1}.$}
\label{tab:Gamma_phonons}
    \centering
    \begin{tabular}{c|c|c|c|c|c|c|c|c}
    \hline \hline
       Mode &\multicolumn{2}{c|}{AF-a \& AF-b} &\multicolumn{2}{c|}{AF-c} &\multicolumn{2}{c|}{Present exp.} &Exp.\cite{pawbake23}  &Exp.\cite{Klein23b} \\
       \hline
       symmetry &Bulk &ML &Bulk &ML &Bulk &ML & & \\
     \hline
    $B_{3g}$ &70.4  &50.4  &70.6  &50.0  &      &   & &71.4 \\
    $B_{2g}$ &70.6  &61.8  &70.7  &63.9  &      &   & &95.0 \\
    $B_{2u}$ &83.1  &81.9  &83.3  &81.7  &      &   & &  \\
    $A_g$    &113.5 &125.9 &113.7 &124.5 &114   &104  &115.0 &121.4 \\
    $B_{3g}$ &160.9 &161.0 &161.0 &164.6 &      &   & &194.7 \\
    $B_{3u}$ &170.1 &169.0 &170.3 &172.1 &      &   & & \\
    $B_{2g}$ &172.8 &169.3 &173.2 &172.4 &      &   & &199.2 \\
    $B_{1u}$ &215.3 &230.4 &215.7 &231.2 &      &   & & \\
    $A_g$    &241.4 &244.5 &241.8 &245.5 &244   &238 &245.5 &254.3 \\
    $B_{3u}$ &287.3 &302.6 &287.6 &308.0 &      &   & & \\
    $B_{2g}$ &297.0 &314.3 &297.3 &318.7 &      &   & &304.8 \\
    $B_{2u}$ &336.9 &356.1 &337.1 &362.6 &      &   & & \\
    $A_g$    &342.2 &363.6 &342.3 &366.5 &343   &341 &344.0 &346.2 \\
    $B_{1u}$ &349.0 &374.1 &349.3 &376.9 & & & & \\
    $B_{3g}$ &364.6 &380.3 &364.5 &386.4 & & & &367.5 \\
    \hline \hline                
    \end{tabular}
\end{table*}
 
\end{document}